\documentclass[twocolumn,showpacs,preprintnumbers,amsmath,amssymb]{revtex4-1}

\usepackage{graphicx}
\usepackage{dcolumn}
\usepackage{bm}

\font\scripti=cmmi7
\font\scriptscripti=cmmi5
\def\sib#1{\setbox0 = \hbox{\scripti #1}
  \kern-.02em\copy0\kern-\wd0
  \kern.04em\box0} 
\def\ssib#1{\setbox0 = \hbox{\scriptscripti #1}
  \kern-.02em\copy0\kern-\wd0
  \kern.04em\box0} 
\font\tenib=cmmib10 
\skewchar\tenib='177 \skewchar\tenib='177 \skewchar\tenib='177
\def\pbold#1{\setbox0 = \hbox{$ #1 $}
  \kern-.022em\copy0\kern-\wd0
  \kern.011em\copy0\kern-\wd0
  \kern.011em\copy0\kern-\wd0
  \kern.011em\copy0\kern-\wd0
  \kern.011em\box0} 

\usepackage{graphicx}
\usepackage{dcolumn}
\usepackage{bm}

\def\lesssim{\ \raise.3ex\hbox{$<$}\kern-0.8em\lower.7ex\hbox{$\sim$}\ }
\def\gesim{\ \raise.3ex\hbox{$>$}\kern-0.8em\lower.7ex\hbox{$\sim$}\ }

\def\diamondcomma{\ \raise.3ex\hbox{$\diamond$}\kern-0.4em\lower.7ex\hbox{$,$}\ }

\def\avg#1{\big <#1 \big >}

\begin{document}
\title{Photoluminescence and gain/absorption spectra of a driven-dissipative electron-hole-photon condensate}
\author{Ryo Hanai}
\email{hanai@acty.phys.sci.osaka-u.ac.jp}
\affiliation{Department of Physics, Osaka University, Toyonaka 560-0043, Japan} 
\affiliation{James Franck Institute and Department of Physics, University of Chicago, Illinois 60637, USA} 

\author{Peter B. Littlewood}
\affiliation{James Franck Institute and Department of Physics, University of Chicago, Illinois 60637, USA} 
\affiliation{Physical Sciences and Engineering Division, Argonne National Laboratory, Argonne, Illinois 60439, USA}

\author{Yoji Ohashi}
\affiliation{Department of Physics, Keio University, Yokohama 223-8522, Japan} 
\date{\today}
\begin{abstract}
We investigate theoretically nonequilibrium effects on photoluminescence and gain/absorption spectra of a driven-dissipative exciton-polariton condensate, by employing the combined Hartree-Fock-Bogoliubov theory with the generalized random phase approximation extended to the Keldysh formalism.
Our calculated photoluminescence spectra is in semiquantitative agreement with experiments, where features such as a blue shift of the emission from the condensate, the appearance of the dispersionless feature of a diffusive Goldstone mode, and the suppression of the dispersive profile of the mode are obtained. 
We show that the nonequilibrium nature of the exciton-polariton condensate strongly suppresses the visibility of the Bogoliubov dispersion in the negative energy branch (ghost branch) in photoluminescence spectra.
We also show that the trace of this branch can be captured as a hole burning effect in gain/absorption spectra. 
Our results indicate that the nonequilibrium nature of the exciton-polariton condensate strongly reduces quantum depletion, while a scattering channel to the ghost branch is still present.
\end{abstract}
\maketitle

\section{Introduction}\label{sec1}
\par

The achievement of the Bose-Einstein condensation (BEC) in an exciton-polariton system \cite{Kasprzak2006} has opened new possibilities to investigate many-body physics in optical devices \cite{Deng2010,Carusotto2013,Byrnes2014,Imamoglu1996}.
Various phenomena analogous to conventional BECs such as a Bogoliubov excitation with a linear dispersion \cite{Utsunomiya2008}, quantum vortices \cite{Lagoudakis2008}, and a non-diffusive transport \cite{Amo2009}, have been observed.
Berezinskii-Kosterlitz-Thouless scaling has also been confirmed recently \cite{Roumpos2012,Nitsche2014,Caputo2016}.
Moreover, at high carrier density, it is expected that one can study an exotic quantum state of matter \cite{Keeling2005,Byrnes2010,Kamide2010,Xue2016} analogous to an ultracold Fermi gas in the Bardeen-Cooper-Schrieffer (BCS)-BEC crossover region \cite{Regal2004,Zwierlein2004,Ohashi2002}. 

\begin{figure}
\begin{center}
\includegraphics[width=0.38\linewidth,keepaspectratio]{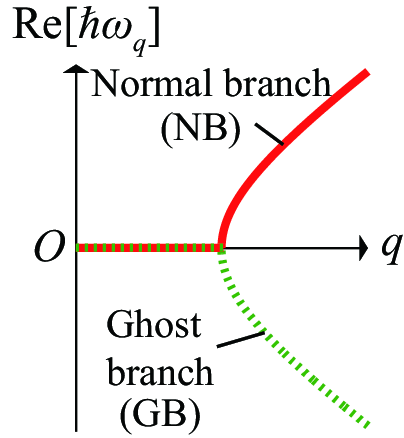}
\end{center}
\caption{(Color online) Real part of the diffusive Goldstone mode (Eq. (\ref{diffNG_intro})). 
At higher momenta, the mode exhibits both the normal (solid line) and the ghost (dotted line) branches, which have positive and negative energy with respect to the energy of the condensate, respectively.}
\label{fig_diffNG}
\end{figure}

\begin{figure*}
\begin{center}
\includegraphics[width=0.7\linewidth,keepaspectratio]{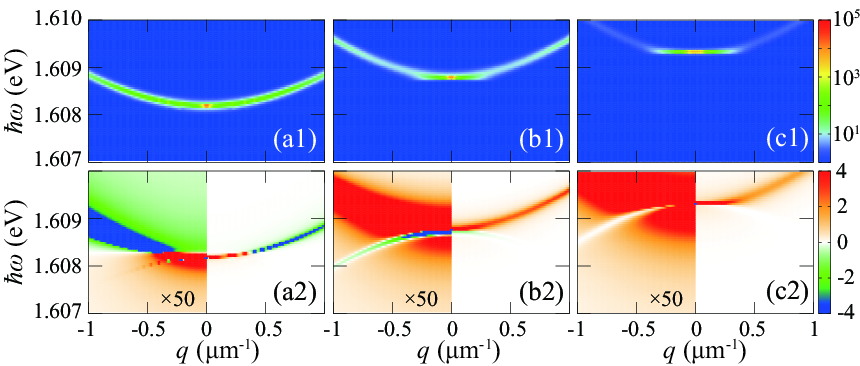}
\end{center}
\caption{(Color online) Calculated pumping dependence of (a1)--(c1) photoluminescence $L(\bm q,\omega)$ and (a2)--(c2) gain/absorption spectra $S(\bm q,\omega)$ on resonance $\delta=0$. The decay rate of cavity photon is set to $\kappa=0.5{\rm meV}$, corresponding to the cavity photon lifetime of $\tau \sim 8~{\rm ps}$. 
[(a1) and (a2)] $\mu_{\rm b}=-4.9 {\rm meV}(\simeq \mu_{\rm b}^{c})$. [(b1) and (b2)] $\mu_{\rm b}=-3.5 {\rm meV}$. [(c1) and (c2)] $\mu_{\rm b}=-2 {\rm meV}$. Parameters are chosen to be as realistic as possible for a GaAs quantum well structure embedded to a microcavity.
The units of the color contour are ${\rm meV}^{-1}$ in all the figures.
For concrete definition of $L(\bm q,\omega),S(\bm q,\omega),\mu_{\rm b},\delta,\gamma$, and $\kappa$, as well as the explicit value of the parameters, see Secs. II and III.
}
\label{figPL}
\end{figure*}

A crucial novelty of the present system is the lossy nature of the optical devices (where the photons in the cavity leak out typically in a timescale of picoseconds) compensated by continuous pumping of the carriers, making the BEC state intrinsically nonequilibrium.
Due to this driven-dissipative nature, a number of novel features are proposed to arise in this system.
For example, the elementary excitation of the condensate is predicted to be the so-called diffusive Goldstone mode \cite{Szymanska2006,Szymanska2007,Wouters2007} ($c$ and $\Gamma$ are real numbers), 
\begin{eqnarray}
\hbar\omega_{\bm q}
= -i\frac{\Gamma}{2}\pm\sqrt{c\bm q^2-\frac{\Gamma^2}{4}},
\label{diffNG_intro}
\end{eqnarray}
with a pure imaginary dispersion $\omega_{\bm q}\propto -i\bm q^2$ at small momenta (see Fig. \ref{fig_diffNG}).
This is in an essential contrast to the equilibrium case $\omega_{\bm q}\propto |\bm q|$.
It has also recently been pointed out that this diffusive character of the mode
makes the nonlinear phase gradient term in the equation of motion to be relevant in a renormalization group sense \cite{Altman2015,Sieberer2016,Wachtel2016}. 
This modifies the Berezinskii-Kosterlitz-Thouless scaling in equilibrium to the Kardar-Parisi-Zhang scaling, which was originally discussed in the context of randomly grown interfaces \cite{Kardar1986}.
In addition, the nonequilibrium feature of the polariton system has brought considerable interest as to how a BEC state of polaritons may evolve into a standard semiconductor laser \cite{Szymanska2006,Szymanska2007,Yamaguchi2012,Yamaguchi2013,Yamaguchi2015,Deng2003,Bajoni2007,Balili2009,Nelsen2009,Kammann2012,Tsotsis2012,
Tempel2012a,Tempel2012b,Horikiri2010,Horikiri2013,Horikiri2016,Horikiri2017,Brodbeck2016}. 

Detailed information on these intrinsic nonequilibrium states is experimentally accessible through the observation of optical properties.
One of the most commonly measured optical quantities is the photoluminescence spectrum (PL).
By detecting the energy as well as its (in-plane) momenta of the leaked-out photons, the distribution of the polaritons can be measured in terms of momentum and energy. 
Using this technique, the dispersionless feature of the diffusive Goldstone mode has been observed \cite{Assmann2011,Tempel2012a,Tempel2012b,Nakayama2016,Nakayama2017,Su2017}.

On the theory side, however, PL has not been understood in a comprehensive way.
Some theoretical studies argue that, in addition to the normal branch (NB) of the diffusive Goldstone mode [solid line in Fig. \ref{fig_diffNG}], the ghost branch (GB) which has negative energy with respect to the condensate energy [dotted line in Fig. \ref{fig_diffNG}] would be populated and appear in PL \cite{Keeling2005,Szymanska2006,Szymanska2007,Marchetti2007,Byrnes2012}.
Especially when a thermal distribution is assumed, the GB was predicted to have the dominant spectral weight at low temperature and high density \cite{Byrnes2012}.
The appearance of the GB can theoretically be attributed to the so-called quantum depletion \cite{Bogoliubov1947}, where particles are kicked out of the condensate due to the repulsive interaction between polaritons. 
Since this effect occurs even in the ground state, the realization of this branch in PL can be regarded as a direct observation of quantum fluctuations of a many-body system.

However, the GB is absent in most of the PL experiments \cite{Kasprzak2006,Utsunomiya2008,Bajoni2007,Nelsen2009,Kammann2012,Tsotsis2012,Tempel2012a,Tempel2012b,Assmann2011,Nakayama2016,Nakayama2017,Su2017,Brodbeck2016}.
There are only a few exceptions that reported observation of GB, which are the experiments done in extremely high density regime \cite{Horikiri2017} 
or in a strongly disordered system \cite{Pieczarka2015} where polaritons were forced to scatter into the GB. 
So far, the suppression of the GB in PL has not been well understood.

On the other hand, a fingerprint of GB has been observed in four-wave mixing experiments \cite{Kohnle2011,Kohnle2012,Wouters2009}.
In such experiments, a pump pulse was resonantly injected to form a polariton condensate, together with a trigger pulse that excites polaritons in the NB at finite momentum.
By the parametric scattering process induced by the third-order nonlinear harmonics of a polariton condensate, the authors have observed a probe four-wave mixing signal in GB \cite{Kohnle2011,Kohnle2012}.
This result implies that GB itself is still present as a scattering channel even in the nonequilibrium situation, while the absence of it in PL implies that the GB is almost unoccupied in the nonequilibrium steady state.

In this paper, by taking into account nonequilibrium effects of the model driven-dissipative electron-hole-photon system, we study the optical properties of the exciton-polariton condensate.
In contrast to prior researches that analyze the Dicke model  (which treats an exciton as a localized excitation)  \cite{Keeling2005,Szymanska2006,Szymanska2007,Marchetti2007} or a Bose gas \cite{Wouters2007,Byrnes2012}, 
we explicitly treat electrons and holes coupled to cavity photons. 
This allows us to safely analyze nonequilibrium effects at density region beyond Mott density and take into account dissociation effects of excitons that plays a crucial role in nonequilibrium states \cite{Hanai2016,Hanai2017,Yamaguchi2012,Yamaguchi2013,Yamaguchi2015}.


Our main result is shown in Fig. \ref{figPL}. (See also Fig. \ref{fig_PL_slice} and Fig. \ref{fig_gain_slice_inset} for tomographic view.)
As seen in Figs. \ref{figPL}(a1)--(c1), we show that the visibility of the GB in PL $L(\bm q,\omega)$ is strongly suppressed by nonequilibrium effects, where a peak structure of the GB is either absent or possess a very small spectral weight.
Our calculated PL is in a qualitative agreement with experiments \cite{Assmann2011,Tempel2012a,Nakayama2016,Nakayama2017,Su2017} 
where it exhibits a blue shift of the emission from the condensate, the appearance of the flat spectrum of the diffusive Goldstone mode, and suppression of spectral weight in the dispersive region at high momentum region, as the pumping power increases.
The amount of the blue shift and the range of the momentum window that exhibits the dispersionless feature have the same order of magnitude as that observed in GaAs experiments \cite{Assmann2011,Tempel2012a}.
Similarly to PL, we show that the GB is also suppressed by nonequilibrium effects in the gain/absorption spectrum $S(\bm q,\omega)$ [Figs. \ref{figPL}(a2)--(c2)], where absorption ($S(\bm q,\omega)>0$) or gain ($S(\bm q,\omega)<0$) from NB is much stronger than that from GB.
However, an optical gain channel from the GB is still present, which appears as either a small but finite gain [Fig. \ref{figPL}(b2)] or suppression of absorption band (hole burning) [Fig. \ref{figPL}(c2)].
The presence of this optical gain channel implies that the GB is still present as a scattering channel.

We also present a systematic analysis on the visibility of the GB, in terms of detuning, pumping power, and the cavity photon decay rate.
We show that, for both in PL and gain/absorption spectra, a clearer visibility of the GB is obtained by setting a bluer detuning or a smaller cavity photon decay rate, that drives the system closer to the equilibrium state.

Our results indicate that the quantum depletion in a driven-dissipative condensate is strongly suppressed by the nonequilibrium nature.
We discuss that this suppression is due to (1) the appearance of the diffusive Goldstone mode, (2) nonequilibrium redistribution of photons, and (3) nonequilibrium-induced pair-breaking effects \cite{Hanai2016,Hanai2017,Yamaguchi2012,Yamaguchi2013} where the dissociated carriers behave as an absorption medium that screens the gain from the GB.

The rest of our paper is organized as follows. 
In Sec. II, we explain our model. 
In Sec. III, we formulate our combined theory of Hartree-Fock-Bogoliubov approximation with the generalized random phase approximation, extended to the Keldysh formalism. 
In Sec. IV, we analyze the PL and gain/absorption spectra of a driven-dissipative electron-hole-photon condensate. In Sec. V, we perform a systematic study on the visibility of the GB, in terms of the pumping power, decay rate of photons, and the detuning. 
In Sec. VI, we give a summary of this paper.

\section{Model driven-dissipative electron-hole-photon system} 

\begin{figure*}
\begin{center}
\includegraphics[width=0.75\linewidth,keepaspectratio]{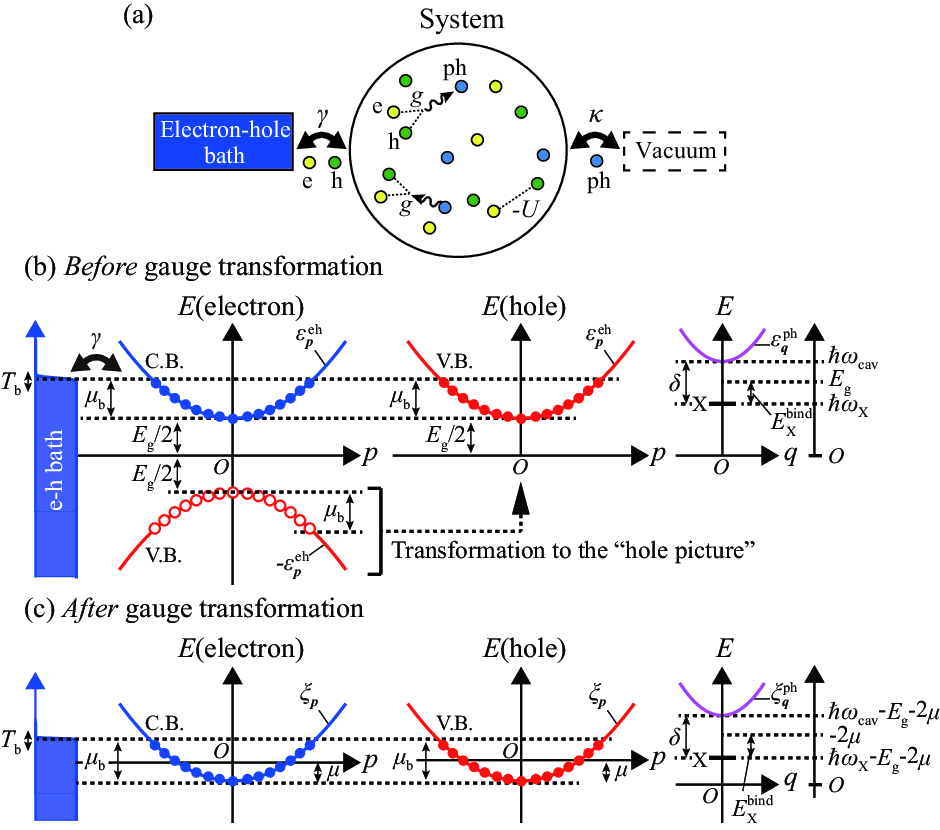}
\end{center}
\caption{(Color online) (a) Model driven-dissipative electron-hole-photon system. 
Electrons and holes are supplied to the system from an electron-hole bath with a thermalization rate $\gamma$. In the system, electrons and holes interact attractively with each other with a coupling constant $-U$, and also pair-annihilate (pair-create) into (from) photons with the coupling constant $g$.
The created photons in the system leak out to a vacuum with a decay rate $\kappa$.
[(b) and (c)] Dispersion of the conduction band (``C.B.''), the valence band (``V.B.''), and the cavity photon, shown in the picture before (b) and after (c) employing the gauge transformation, described in the paragraph below Eq. (\ref{Ep}).
In the left panel of (b), the valence band is described in the ``electron picture'', while in the center panel of (b) and (c), it is described in a ``hole picture''.
Here, the bath drives the electron-hole carriers to the Fermi distribution  (Eq. (\ref{fb})), characterized by the bath-chemical-potential $\mu_{\rm b}$ and the temperature $T_{\rm b}$.
In the right panel, photon dispersion, as well as an exciton state ``X'' which lies at  $\hbar\omega_{\rm X}=E_{\rm g}-E_{\rm X}^{\rm bind}$, are schematically shown.
The detuning $\delta$ is defined as the energy difference between the photon energy at $\bm q=0$ and the exciton energy, i.e., $\delta=\hbar\omega_{\rm cav}-\hbar\omega_{\rm X}$.
}
\label{figmodel}
\end{figure*}

The model driven-dissipative electron-hole-photon system we consider in this paper is shown schematically in Fig. \ref{figmodel}(a) \cite{Szymanska2006,Szymanska2007,Yamaguchi2012,Yamaguchi2013,Yamaguchi2015},
which consists of the system, an electron-hole bath, and a vacuum.
In this model, the photon pumping of carriers, as well as its thermalization is modeled as an attachment of an electron-hole bath to the system. 
The bath continuously injects electron-hole carriers to the system and thermalize them with a thermalization rate $\gamma$. 
Then, the injected carriers in the system pair-annihilate (or pair-create) to photons.
The created photons leak out to the vacuum with a decay rate $\kappa$, driving the system into a nonequilibrium steady state.

The above model is described by the Hamiltonian, $H=H_{\rm s}+H_{\rm t}+H_{\rm env}$. 
$H_{\rm s}$ describes the relevant system consisting of electrons in the conduction band, holes in the valence band, and cavity photons, given by
\begin{eqnarray}
H_{\rm s}&=&\sum_{\bm p,\sigma}\varepsilon^{\rm eh}_{\bm p}c^\dagger_{\bm p\sigma}c_{\bm p\sigma}
+\sum_{\bm q}\varepsilon_{\bm q}^{\rm ph}a^\dagger_{\bm q}a_{\bm q}
\nonumber\\
&-&U
\sum_{\bm p,\bm p',\bm q}c^\dagger_{\bm p+\bm q/2,{\rm e}}c^\dagger_{-\bm p+\bm q/2,{\rm h}}c_{-\bm p'+\bm q/2,{\rm h}}c_{\bm p'+\bm q/2,{\rm e}}
\nonumber\\
&+&g\sum_{\bm p,\bm q}
[a^\dagger_{\bm q}c_{-\bm p+\bm q/2,{\rm h}}c_{\bm p+\bm q/2,{\rm e}}+{\rm h.c.}].
\label{Hs}
\end{eqnarray}
Here, $c_{{\bm p},\sigma={\rm e(h)}}$ and $a_{\bm q}$ are an annihilation operator of an electron (hole) and a cavity photon, respectively. 
The first term in Eq. (\ref{Hs}) is the kinetic term of the electrons and holes, which is assumed to have the same mass $m_{\rm eh}$ and the kinetic energy $\varepsilon^{\rm eh}_{\bm p}=\hbar^2 {\bm p}^2/(2m_{\rm eh})+E_{\rm g}/2$ ($E_{\rm g}$ is an energy gap of the semiconductor quantum well).
Here, we have transformed the empty states of the valence band in the ``electron picture'' shown in the left panel of Fig. \ref{figmodel}(b) (where the kinetic energy of electrons in the valence band is given by $-\varepsilon_{\bm p}^{\rm eh}$)
to the valence band holes shown in the center panel of Fig. \ref{figmodel}(b) (where the kinetic energy of holes in the valence band is given by $\varepsilon_{\bm p}^{\rm eh}$).
We also briefly note that $\sigma={\rm e,h}$ represents the electron and hole component in our notation (not the spins of the band), and have neglected the spin degrees of freedom.

The second term describes the kinetic term of cavity photons with the kinetic energy $\varepsilon^{\rm ph}_{\bm q}=\hbar\omega_{\rm cav}+\hbar^2 {\bm q}^2/(2m_{\rm ph})$ (schematically shown in the right panel of Fig. \ref{figmodel}(b)). Here, $m_{\rm ph}=n_{\rm c}^2\hbar /c (2\pi/\lambda)$ is a cavity photon mass, and $\hbar\omega_{\rm cav}=(c/n_{\rm c})\hbar (2\pi/\lambda)$ is assumed to be controllable by varing the microcavity length $\lambda$ ($n_{\rm c}$ is the reflactive index of the cavity).

The third term describes an attractive interaction between electrons and holes, 
and the fourth term describes the dipole coupling between electron-hole carriers and photons within the rotation wave approximation.
Here, the interaction between electrons and holes is assumed to be an attractive contact-type interaction $-U<0$, instead of a realistic long-range Coulomb interaction.
We expect that this simplication will not affect the low energy excitation properties, at least in a qualitative way, since for both the cases, the gapless mode is present, attributed to the neutrality of the electron-hole-photon condensate where the Anderson-Higgs mechanism is absent \cite{Cote1988}.
$g$ describes the dipole-coupling strength between the electron-hole carriers and photons.

For latter use, it is useful to point out that, in the absence of cavity photons, an electron and a hole forms an exciton in the dilute limit.
The energy level of this state lies at $\hbar\omega_{\rm X}=E_{\rm g}-E_{\rm X}^{\rm bind}$ (``X'' in the right panel of Fig. \ref{figmodel}(b)), where $E_{\rm X}^{\rm bind}$ is the exciton binding energy.
This allows us to define the detuning parameter between the cavity photon and the exciton state, (See the right panel of Fig. \ref{figmodel}(b).)
\begin{eqnarray}
\delta=\hbar\omega_{\rm cav}-\hbar\omega_{\rm X}.
\end{eqnarray}

Although the formalism we develop in this paper can be applied to any Wannier-type polariton systems, in this paper, we set the parameters to be as realistic as possible for a GaAs quantum well structure embedded in a microcavity. 
We set $m_{\rm eh} = 0.068m_0,m_{\rm ph} = 3\times 10^{-5}m_0$ (where $m_0$ is the electron mass), $E_{\rm g}=1.624 {\rm eV}$, and the cutoff wave number $k_{\rm c}=2\pi / a=1360 {\rm \mu m}^{-1}$ (where $a=4.6{\rm nm}$ is the lattice constant). 

The magnitude of $U$ and $g$ are determined to reproduce the measured exciton and polariton energy level in the GaAs quantum well structure.
As explained in Appendix \ref{App_parameter}, we have chosen $U=5.2 {\rm meV / \mu m^2}$ in order to reproduce the exciton spectrum to lie at $\hbar\omega_{\rm X}=1.614{\rm eV}$ in the dilute limit, where $E_{\rm X}^{\rm bind}=10{\rm meV}$.
Similarly, we have chosen the dipole coupling constant as $g=1.7{\rm meV/\mu m^2}$ to reproduce the Rabi splitting of $2g_{\rm R}=14{\rm meV}$ in the dilute limit on resonance $\delta = 0$.
With these choices of parameters, we have checked that the lower polariton energy level lie approximately at $\hbar\omega_{\rm LP}=[\hbar\omega_{\rm cav}+\hbar\omega_{\rm X}- \sqrt{\delta^2+4g_{\rm R}^2}]/2$ (See Fig. \ref{fig_ELP} in Appendix \ref{App_parameter}.), consistent with the conventional polariton picture \cite{Deng2010}.

Electrons and holes are incoherently supplied to the system via a tunneling $\Gamma_{\rm b}$ from an electon-hole bath, while cavity photons decay to vacuum via a tunneling $\Gamma_{\rm v}$. 
These processes are described by the tunneling Hamiltonian, 
\begin{eqnarray}
H_{\rm t}&=&
\sum_{\bm p,\bm P,\sigma,i}[\Gamma_{\rm b}c^\dagger_{\bm p,\sigma}b_{\bm P,\sigma}e^{i\bm p\cdot\bm r_i}e^{-i\bm P\cdot \bm R_i}+{\rm h.c.}]
\nonumber\\
&+&
\sum_{\bm q,\bm Q,i}[\Gamma_{\rm v}a^\dagger_{\bm q}\psi_{\bm Q}e^{i\bm q\cdot\bm r_i}e^{-i\bm Q\cdot \bm R_i}+{\rm h.c.}], 
\label{Ht}
\end{eqnarray}
where the bath and the vacuum are described by the Hamiltonian,
\begin{eqnarray}
H_{\rm env}&=&\sum_{\bm P,\sigma}\varepsilon_{\bm P}^{\rm eh, b}b^\dagger_{\bm P,\sigma}b_{\bm P,\sigma}+\sum_{\bm Q}\varepsilon_{\bm Q}^{\rm ph,v}\psi^\dagger_{\bm Q}\psi_{\bm Q}.
\label{Henv}
\end{eqnarray}
Here, $b_{\bm P,{\rm e(h)}}$ is an annihilation operator of a bath electron (hole), and $\psi_{\bm Q}$ is an annihilation operator of a vaccum photon.
$\varepsilon^{\rm eh,b}_{\bm P,{\rm e(h)}}$ and $\varepsilon^{\rm ph,v}_{\bm Q}$ are the kinetic energy of bath electrons (holes) and vacuum photons, respectively.
We have assumed in Eq. (\ref{Ht}) that the particles tunnel from random positions $\bm r_i$ in the system to $\bm R_i$ in the bath or vacuum ($i=1,2,...,N_{\rm t}$) \cite{NOTErandom}.
As derived in Appendix \ref{App_HFB} within the second-order Born approximation, these couplings to the bath and the vacuum induce the thermalization rate of electron-hole carriers,
\begin{eqnarray}
\gamma=\pi N_{\rm t}\rho_{\rm b}|\Gamma_{\rm b}|^2,
\label{gamma}
\end{eqnarray}
as well as the decay rate to the vacuum, 
\begin{eqnarray}
\kappa=\pi N_{\rm t}\rho_{\rm v}|\Gamma_{\rm v}|^2,
\label{kappa}
\end{eqnarray}
where $\rho_{\rm b(v)}$ is the bath (vacuum) density of states assumed to be constant.
In this paper, we set the thermalization rate to $\gamma=4{\rm meV}$.

The bath and the vacuum are assumed to be large compared to the system, and they stay in an equilibrium state. The bath distribution is given by
\begin{eqnarray}
f_{\rm b}(\omega)=\frac{1}{\exp[(\hbar\omega-(\mu_{\rm b}+E_{\rm g}/2))/(k_{\rm B}T_{\rm b})]+1},
\label{fb}
\end{eqnarray}
characterized by its chemical potential $\mu_{\rm b}$ and temperature $T_{\rm b}$
(see Fig. \ref{figmodel}(b)).
The electron-hole density monotonically increases as $\mu_{\rm b}$ increases; thus, $\mu_{\rm b}$ corresponds to the pumping power in our model. 
The vacuum distribution is assumed to vanish, i.e., $b_{\rm v}(\omega)=0$.

In this paper, we analyze steady-state properties of the Bose-condensed phase of the above model, characterized by the order parameter \cite{Holland2001,Ohashi2002,Ohashi2003}, 
\begin{eqnarray}
\Delta(t)=U\sum_{\bm p}\avg{c_{-\bm p,{\rm h}}c_{\bm p,{\rm e}}}-g\avg{a_0}.
\label{Delta}
\end{eqnarray}
We employ an steady-state ansatz \cite{Szymanska2006,Szymanska2007,Yamaguchi2012,Yamaguchi2013,Yamaguchi2015,Hanai2016,Hanai2017}
\begin{eqnarray}
\Delta(t)=\Delta_0 e^{-i(2\mu+E_{\rm g}) t/\hbar},
\label{Delta0}
\end{eqnarray}
where we have introduced a parameter $\mu$ which works as the ``chemical potential'' of the system.
With our choice of the order parameter (Eq. (\ref{Delta})), a single-particle excitation energy is given by a conventional form, 
\begin{eqnarray}
E_{\bm p}=\sqrt{(\varepsilon_{\bm p}-\mu)^2+\Delta_0^2},
\label{Ep}
\end{eqnarray}
as derived in Appendix \ref{App_HFB}.

In explicit calculations we perform below, it is convenient to employ the gauge transformation $c_{\bm p,\sigma}\rightarrow e^{i(\mu+E_{\rm g}/2) t/\hbar}c_{\bm p,\sigma},b_{\bm P,\sigma}\rightarrow e^{i(\mu+E_{\rm g}/2) t/\hbar}b_{\bm P,\sigma},a_{\bm q}\rightarrow e^{i(2\mu + E_{\rm g})t/\hbar}a_{\bm q}$, and $\psi_{\bm Q}\rightarrow e^{i(2\mu + E_{\rm g})t/\hbar}\psi_{\bm Q}$ in order to formally eliminate the time dependence of the order parameter in Eq. (\ref{Delta}) \cite{Szymanska2006,Szymanska2007,Yamaguchi2012,Yamaguchi2013,Yamaguchi2015,Hanai2016,Hanai2017}.
In practice, this transformation is performed by replacing $\varepsilon^{\rm eh}_{\bm p},\varepsilon_{\bm q}^{\rm eh,b},\varepsilon^{\rm ph}_{\bm q},\varepsilon_{\bm Q}^{\rm v}$, and $\mu_{\rm b}$ by $\xi_{\bm p}=\varepsilon^{\rm eh}_{\bm p} - \mu-E_{\rm g}/2, \xi_{\bm P}^{\rm b}=\varepsilon^{\rm eh,b}_{\bm P}-\mu -E_{\rm g}/2, \xi_{\bm q}^{\rm ph}=\varepsilon_{\bm q}^{\rm ph}-2\mu-E_{\rm g}, \varepsilon_{\bm Q}^{\rm v,ph}-2\mu-E_{\rm g}$, and $\mu_{\rm b}-\mu$, respectively. 
By this transformation, the origin of the cavity photon energy $\hbar\omega$ is also shifted as $\hbar\omega'=\hbar\omega-2\mu-E_{\rm g}$.
The resulting energy levels are schematically described in Fig. \ref{figmodel}(c).

\section{Generalized random phase approximation} \label{GRPA}

We now develop a generalized random phase approximation (GRPA) combined with the Hartree-Fock-Bogoliubov (HFB) theory extended to the Keldysh formalism, to analyze photoluminescence (PL) and gain/absorption spectra of an exciton-polariton condensate in a nonequilibrium steady state. 
In this approach, we first determine the nonequilibrium steady state of the Bose-condensate within the HFB-Keldysh theory, and then compute the fluctuations around that steady state to obtain the optical properties.
The HFB-Keldysh theory of the above model (Eqs. (\ref{Hs})--(\ref{Henv})) has been shown \cite{Yamaguchi2012,Yamaguchi2013,Yamaguchi2015} to capture the essential features of the BCS-BEC crossover \cite{Byrnes2010,Kamide2010,Xue2016} and their connection to a conventional semiconductor laser.
Since our GRPA formalism treats fluctuations in a fully consistent manner to the HFB-Keldysh theory, we can safely analyze optical properties in this BEC-BCS-laser crossover context.

The steady-state solution within the HFB-Keldysh theory can be obtained by solving the steady-state gap equation, given by \cite{Yamaguchi2012,Yamaguchi2013,Yamaguchi2015,Hanai2016,Hanai2017}, 
(For derivation, see Appendix \ref{App_HFB}.)
\begin{eqnarray}
\frac{1}{U_{\rm eff}}= \sum_{\bm p}\int \frac{\hbar d\omega} {\pi}
\frac{
F_{-}(\omega)\hbar\omega
+
F_{+}(\omega)[\xi_{\bm p}+i\gamma]
}
{[(\hbar\omega - E_{\bm p})^2 + \gamma^2]
[(\hbar\omega + E_{\bm p})^2 + \gamma^2]
}.
\nonumber\\
\label{GAP}
\end{eqnarray}
Here, 
$F_{\rm \pm}(\omega)=[F(\omega)\pm F(-\omega)]/2$ and $F(\omega)=\gamma[1-2f_{\rm b}(\omega)].$
\begin{eqnarray}
U_{\rm eff}=U+\frac{g^2}{\hbar\omega_{\rm cav}-2\mu-E_{\rm g}-i\kappa}
\label{Ueff}
\end{eqnarray}
describes the effective interaction between the electrons and holes, where the first term is a bare electron-hole interaction, and the second arises from a second-order process of photon emission and absorption.
From the (complex) gap equation (\ref{GAP}), we determine the order parameter $\Delta_0$ and $\mu$.

Once all the parameter sets of the steady state ($\Delta_0,\mu$) are determined by solving Eq. (\ref{GAP}), we can move on to the analysis of optical properties.
In exciton-polariton experiments, PL is measured by detecting the energy of the leaked-out photons from the cavity in an angle-resolved way.
Strictly speaking, the reflectance of the Bragg mirror which determines the decay rate of the photons from the microcavity has an angle $\theta(=\arctan(|\bm q|/(2\pi\lambda))$ dependence, which however, are negligibly small as long as we consider small angle $\theta\ll 1$.
Similarly, the energy dependence of cavity photons on the leakage rate is also negligible in the energy region that is measured \cite{Deng2010}.
As a result, the intensity of PL is nearly proportional to the occupied spectral weight function of the cavity photons, 
\begin{eqnarray}
L(\bm q,\omega)=\frac{1}{2\pi}\int_{-\infty}^\infty d(t-t') e^{-i\omega (t-t')} \avg{a^\dagger_{\bm q}(t')a_{\bm q}(t)}.
\nonumber\\
\label{PL}
\end{eqnarray}
We simply call $L(\bm q,\omega)$ ``PL'' in this paper.
This quantity can be obtained by computing the Nambu-Keldysh Green's function of the cavity photons in a steady state,
\begin{widetext}
\begin{eqnarray}
\hat D(\bm q,t-t')
&=&
\left(
\begin{array}{cc}
\hat D_{aa}(\bm q,t-t') & \hat D_{ab}(\bm q,t-t') \\
\hat D_{ba}(\bm q,t-t') & \hat D_{bb}(\bm q,t-t')
\end{array}
\right)=
\left(
\begin{array}{cc}
\hat D^{\rm R}(\bm q,t-t') & \hat D^{\rm K}(\bm q,t-t') \\
0                                 & \hat D^{\rm A}(\bm q,t-t')
\end{array}
\right)
\nonumber\\
&=&
-i\left(
\begin{array}{cc}
\theta(t-t')\avg{[\hat A_{\bm q}(t)\diamondcomma \hat A^\dagger_{\bm q}(t')]} 
& \avg{\hat A_{\bm q}(t)\diamond \hat A^\dagger_{\bm q}(t') + \hat A^\dagger_{\bm q}(t')\diamond \hat A_{\bm q}(t)}\\
0 
& \theta(t'-t)\avg{[\hat A_{\bm q}(t)\diamondcomma \hat A^\dagger_{\bm q}(t')]}
\end{array}
\right).
\label{Dph}
\end{eqnarray}
Here, we have introduced a Nambu field of cavity photons,
\begin{eqnarray}
\hat A_{\bm q} = 
\left(
\begin{array}{c}
a_{\bm q} \\
a_{-\bm q}^\dagger
\end{array}
\right)
\equiv 
\left(
\begin{array}{c}
A_{\bm q,1} \\
A_{\bm q,2}
\end{array}
\right),
\end{eqnarray}
and
\begin{eqnarray}
\big(
\hat A_{\bm q}(t)\diamond \hat A_{\bm q}^\dagger(t')
\big)_{s,s'}
&\equiv&
A_{\bm q,s}(t)A_{\bm q,s'}^\dagger(t')
=
\left(
\begin{array}{cc}
a_{\bm q}(t)             a_{\bm q}^\dagger(t')  & a_{\bm q}(t)              a_{-\bm q}(t')\\
a_{-\bm q}^\dagger(t) a_{\bm q}^\dagger(t') & a_{-\bm q}^\dagger(t) a_{-\bm q}(t')
\end{array}
\right)_{s,s'}
,\\
\big(
\hat A^\dagger_{\bm q}(t')\diamond \hat A_{\bm q}(t)
\big)_{s,s'}
&\equiv&
A^\dagger_{\bm q,s'}(t')A_{\bm q,s}(t)
=\left(
\begin{array}{cc}
a^\dagger_{\bm q}(t')  a_{\bm q}(t)             & a_{-\bm q}(t') a_{\bm q}(t)\\
a_{\bm q}^\dagger(t') a_{-\bm q}^\dagger(t) & a_{-\bm q}(t') a_{-\bm q}^\dagger(t) 
\end{array}
\right)_{s,s'}.
\end{eqnarray}
\end{widetext}
PL can be obtained by taking the (1,1)-component in Nambu space of the lesser component $\hat D^<=[-\hat D^{\rm R}+\hat D^{\rm A}+\hat D^{\rm K}]/2$,
\begin{eqnarray}
L(\bm q,\omega)
&=&
\frac{i}{2\pi}\int_{-\infty}^\infty d(t-t') e^{i\omega (t-t')} D^{<}_{11}(\bm q,t-t')
\nonumber\\
&=&
\frac{i}{2\pi}D^{<}_{11}(\bm q,\omega).
\end{eqnarray}
One can also compute the gain/absorption spectrum $S(\bm q,\omega)$ from Eq. (\ref{Dph}), as 
\begin{eqnarray}
S(\bm q,\omega)
&=&-\frac{1}{\pi}{\rm Im}\int_{0}^\infty d(t-t') e^{i\omega (t-t')} \avg{[a_{\bm q}(t),a^\dagger_{\bm q}(t')]}
\nonumber\\
&=&-\frac{1}{\pi}\int_{-\infty}^\infty d(t-t') e^{i\omega (t-t')} {\rm Im}D^{\rm R}_{11}(\bm q,t-t')
\nonumber\\
&=&
-\frac{1}{\pi}{\rm Im} D^{\rm R}_{11}(\bm q,\omega).
\label{SW}
\end{eqnarray}
which describes  absorption ($S(\bm q,\omega)>0$) or  gain ($S(\bm q,\omega)<0$) of photons in the cavity.

\begin{figure*}
\begin{center}
\includegraphics[width=0.75\linewidth,keepaspectratio]{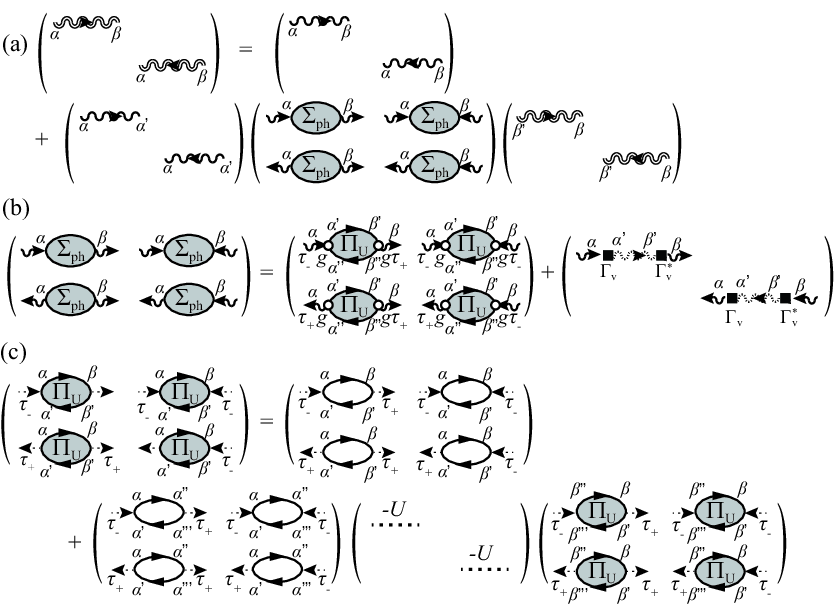}
\end{center}
\caption{(Color online) Diagrammatic expression of the dressed Green's function $\hat D$. (a) Dyson's equation for the dressed photon Green's function $\hat D$. (b) Photon self-energy $\hat\Sigma_{\rm ph}$. (c) Pairing fluctuations $\hat\Pi_{U}$ induced by coupling $-U$. Here, the double and the single wavy line denote the dressed photon Green's function $\hat D$ and $\hat D^0$, respectively.
The solid line describes the electron-hole single-particle Green's function in the nonequilibrium steady state $\hat G$. The dotted line and the open circle represent the electron-hole coupling $-U$ and photon-electron-hole dipole coupling $g$, respectively. The dashed-wavy line represents $\hat B_{\rm v}$, and the solid rectangle describes the tunneling $\Gamma_{\rm v}$. Greek indices $\alpha, \beta, {\rm etc.} (= +,-)$ denotes the indices of the Keldysh space.}
\label{figdiagrambubble}
\end{figure*}

From below, for convenience, we employ the gauge transformation described in the paragraph below Eq. (\ref{Ep}).
To distinguish the gauge-transformed quantities from the quantities written in the original picture, we denote them by putting ``$~\bar{}~$''.
For example, we denote the PL in the gauge transformed picture as $\bar L(\bm q,\omega)$, which is related to the PL in the original picture $L(\bm q,\omega)$ as,
(See Fig. \ref{figmodel}(c).)
\begin{eqnarray}
\bar L(\bm q,\omega+2\mu+E_{\rm g})=L(\bm q, \omega).
\label{Lbar}
\end{eqnarray}

PL and gain/absorption spectra are determined by fluctuations of the condensate around the steady state. To compute them, they must be treated in a consistent manner with the approximation employed in the computation of the steady state, otherwise they violate the gauge invariance of the system (Ward's identity \cite{Schrieffer}).
This becomes especially important when discussing collective excitations of the steady state, since the gauge invariance is directly related to the appearance of a gapless mode \cite{Szymanska2006,Szymanska2007,Hanai2017} (the Goldstone's theorem \cite{Goldstone1961}).
In our HFB-Keldysh case, Fig. \ref{figdiagrambubble} summarizes the diagrams that satisfies the above demand. 
(An anologous case is studied for equilibrium case in Ref. \cite{Ohashi2003}, in the context of an ultracold Fermi gas.)
Here, Fig. \ref{figdiagrambubble}(a) describes the Dyson's equation for the (gauge-transformed) cavity photon Green's function $\hat {\bar D}$,
\begin{eqnarray}
\hat {\bar D}_{\alpha,\beta}(\bm q,\omega)
&=&
\hat {\bar D}^0_{\alpha,\beta}(\bm q,\omega)
\nonumber\\
&+&\hat {\bar D}^0_{\alpha,\alpha'}(\bm q,\omega)\hat{\bar \Sigma}^{\rm ph}_{\alpha',\beta'}(\bm q,\omega)\hat {\bar D}_{\beta',\beta}(\bm q,\omega),
\label{Dysonph}
\end{eqnarray}
where a free Green's function of the cavity photons is given by,
\begin{widetext}
\begin{eqnarray}
\hat {\bar D}^0(\bm q,\omega)
&=&
\left(
\begin{array}{cc}
\hat {\bar D}^0_{aa}(\bm q,\omega) & \hat {\bar D}^0_{ab}(\bm q,\omega) \\
\hat {\bar D}^0_{ba}(\bm q,\omega) & \hat {\bar D}^0_{bb}(\bm q,\omega)
\end{array}
\right)=
\left(
\begin{array}{cc}
\hat {\bar D}^{0{\rm R}}(\bm q,\omega) & \hat {\bar D}^{0{\rm K}}(\bm q,\omega) \\
0                                                 & \hat {\bar D}^{0{\rm A}}(\bm q,\omega)
\end{array}
\right)
\nonumber\\
&=&
\left(
\begin{array}{cc}
[(\hbar\omega+i\delta)\tau_3 -\xi_{\bm q}^{\rm ph}]^{-1} & -\pi i (1+2b_{\rm ph}(\omega))\delta(\hbar\omega\tau_3-\xi_{\bm q}^{\rm ph}) \\
0                  & [(\hbar\omega-i\delta)\tau_3 -\xi_{\bm q}^{\rm ph}]^{-1}
\end{array}
\right).
\end{eqnarray}
Here, $\tau_i (i=1,2,3)$ are the Pauli matrices acting on the Nambu space.
Equation (\ref{Dysonph}) can be formally solved as, 
\begin{eqnarray}
\hat {\bar D}^{\rm R}(\bm q,\omega)
&=&
\big[
[\hat {\bar D}^{0{\rm R}}(\bm q,\omega)]^{-1}-\hat {\bar \Sigma}_{\rm ph}^{\rm R}(\bm q,\omega)
\big]^{-1},
\\
\hat {\bar D}^{\rm A}(\bm q,\omega)
&=&
\big[
[\hat {\bar D}^{0{\rm A}}(\bm q,\omega)]^{-1}-\hat {\bar \Sigma}_{\rm ph}^{\rm A}(\bm q,\omega)
\big]^{-1}
(= [\hat {\bar D}^{\rm R}(\bm q,\omega)]^\dagger),\\
\hat {\bar D}^{\rm K}(\bm q,\omega)
&=&
\hat {\bar D}^{\rm R}(\bm q,\omega)
\hat {\bar \Sigma}_{\rm ph}^{\rm K}(\bm q,\omega)
\hat {\bar D}^{\rm A}(\bm q,\omega)
+\big[
1+\hat {\bar D}^{\rm R}(\bm q,\omega)\hat {\bar \Sigma}_{\rm ph}^{\rm R}(\bm q,\omega)
\big] 
\hat {\bar D}^{0{\rm K}}(\bm q,\omega)
\big[
1+\hat {\bar \Sigma}_{\rm ph}^{\rm A}(\bm q,\omega)\hat {\bar D}^{\rm A}(\bm q,\omega)
\big].
\label{DK}
\end{eqnarray}
\end{widetext}
Here, $b_{\rm ph}(\omega)$ in the Keldysh component of $\hat {\bar D}^0$ is the initial distribution of photons, which however, does not affect the final result, since the second term in Eq. (\ref{DK}) can be shown to vanish as long as ${\rm Im}\Sigma^{\rm R}_{\rm ph}(\omega)\ne 0$ for arbitrary $\omega$.
The self-energy for cavity photons
\begin{eqnarray}
\hat {\bar \Sigma}_{\rm ph}(\bm q,\omega)
&=&
\left(
\begin{array}{cc}
\hat {\bar \Sigma}^{\rm ph}_{aa}(\bm q,\omega) & \hat {\bar \Sigma}^{\rm ph}_{ab}(\bm q,\omega) \\
\hat {\bar \Sigma}^{\rm ph}_{ba}(\bm q,\omega) & \hat {\bar \Sigma}^{\rm ph}_{bb}(\bm q,\omega)
\end{array}
\right)
\nonumber\\
&=&
\left(
\begin{array}{cc}
\hat {\bar \Sigma}_{\rm ph}^{\rm R}(\bm q,\omega) & \hat {\bar \Sigma}_{\rm ph}^{\rm K}(\bm q,\omega) \\
0                                              & \hat {\bar \Sigma}_{\rm ph}^{\rm A}(\bm q,\omega)
\end{array}
\right),
\end{eqnarray}
is given diagramatically in Fig. \ref{figdiagrambubble}(b), where its explicit form is given by,
\begin{eqnarray}
\hat{\bar \Sigma}^{\rm ph}_{\alpha,\beta}(\bm q,\omega)
&=&
ig^2\tilde\gamma^\alpha_{\alpha',\alpha''} 
[\hat{\bar\Pi}_{U}(\bm q,\omega)]^{\alpha'\beta'}_{\alpha'',\beta''}
\gamma^{\beta}_{\beta',\beta''}
\nonumber\\
&+&
N_{\rm t}|\Gamma_{\rm v}|^2\sum_{\bm Q}\hat {\bar B}^{\rm v}_{\alpha,\beta}(\bm Q,\omega).
\label{Sigph}
\end{eqnarray}
The first term of Eq. (\ref{Sigph}) describes the dipole coupling of photons to the electron-hole pairing fluctuations $\hat{\bar \Pi}_{U}$, induced by electron-hole coupling $-U$ (Fig. \ref{figdiagrambubble}(c)), given by,
\begin{eqnarray}
&&[\hat{\bar \Pi}_{U}(\bm q,\omega)]^{\alpha,\beta}_{\alpha',\beta'}
=[\hat{\bar \Pi}(\bm q,\omega)]^{\alpha,\beta}_{\alpha',\beta'}
\nonumber\\
&&+i(-U)
[\hat{\bar \Pi}(\bm q,\omega)]^{\alpha,\alpha''}_{\alpha',\alpha'''}
\eta^{\alpha'',\beta''}_{\alpha''',\beta'''}
[\hat{\bar \Pi}_{U}(\bm q,\omega)]^{\beta'',\beta}_{\beta''',\beta'}.
\label{PiU_index}
\end{eqnarray}
Here, $\eta^{\alpha,\alpha'}_{\beta,\beta'},\gamma_{\alpha,\beta}^{\beta'}$, and $\tilde\gamma_{\alpha,\beta}^{\beta'}$ are vertices at an electron-hole coupling, emission and absorption of photons, respectively, with the form \cite{Rammer}
\begin{eqnarray}
\eta^{\alpha,\alpha'}_{\beta,\beta'}&=&\frac{1}{2}(\delta_{\alpha,\beta}\delta_{\alpha',-\beta'}+\delta_{\alpha,-\beta}\delta_{\alpha',\beta'}),\\
\gamma_{\alpha,\beta}^+&=&\tilde\gamma_{\alpha,\beta}^-=\frac{1}{\sqrt{2}}\delta_{\alpha,\beta} \\
\gamma_{\alpha,\beta}^-&=&\tilde\gamma_{\alpha,\beta}^+=\frac{1}{\sqrt{2}}\delta_{\alpha,-\beta}.
\end{eqnarray} 
$\hat{\bar \Pi}$ in Eq. (\ref{PiU_index}) is the lowest-order pair correlation function, given by,
\begin{eqnarray}
[\hat{\bar \Pi}(\bm q,\omega)]^{\alpha,\beta}_{\alpha',\beta'}
=
\left(
\begin{array}{cc}
[{\bar \Pi}_{-+}(\bm q,\omega)]^{\alpha,\beta}_{\alpha',\beta'} & [{\bar \Pi}_{--}(\bm q,\omega)]^{\alpha,\beta}_{\alpha',\beta'} \\
{[}{\bar \Pi}_{++}(\bm q,\omega)]^{\alpha,\beta}_{\alpha',\beta'} & [{\bar \Pi}_{+-}(\bm q,\omega)]^{\alpha,\beta}_{\alpha',\beta'} 
\end{array}
\right),
\nonumber\\
\end{eqnarray}
where
\begin{eqnarray}
[{\bar \Pi}_{s,s'}(\bm q,\omega)]^{\alpha,\beta}_{\alpha',\beta'}
&=&
-\sum_{\bm k}\int_{-\infty}^{\infty}\frac{\hbar d\omega_1}{2\pi}
\nonumber\\
&\times&
{\rm Tr} \Big[
\tau_s \hat{\bar G}_{\alpha,\beta}(\bm k+\frac{\bm q}{2},\omega_1+\frac{\omega}{2}) 
\nonumber\\
&\times&
\tau_{s'}  \hat{\bar G}_{\beta',\alpha'}(\bm k-\frac{\bm q}{2},\omega_1-\frac{\omega}{2})
\Big].
\label{PI}
\end{eqnarray}
Here, $\hat \Pi$ is calculated using the electron-hole Nambu-Keldysh single-particle Green's function in the steady state $\hat {\bar G}$ obtained within the HFB-Keldysh theory. 
Their explicit form, as well as their derivation is given in Appendix \ref{App_HFB}.

\begin{figure}
\begin{center}
\includegraphics[width=0.6\linewidth,keepaspectratio]{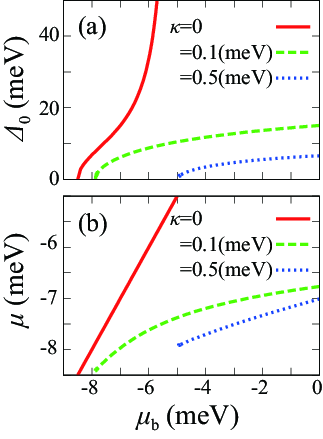}
\end{center}
\caption{(Color online) Calculated steady-state solution on resonance $\delta=0$, as a function of the pumping power $\mu_{\rm b}$. (a) Order parameter $\Delta_0$. (b) System ``chemical potential'' $\mu$.}
\label{fig_delta_mu}
\end{figure}

The second term in Eq. (\ref{Sigph}) describes the effects of tunneling to the vacuum within the second-order Born approximation, where
\begin{widetext}
\begin{eqnarray}
\hat {\bar B}_{\rm v}(\bm Q,\omega)
&=&
\left(
\begin{array}{cc}
\hat {\bar B}^{\rm v}_{aa}(\bm Q,\omega) & \hat {\bar B}^{\rm v}_{ab}(\bm Q,\omega) \\
\hat {\bar B}^{\rm v}_{ba}(\bm Q,\omega) & \hat {\bar B}^{\rm v}_{bb}(\bm Q,\omega)
\end{array}
\right)=
\left(
\begin{array}{cc}
\hat {\bar B}^{\rm R}_{\rm v}(\bm Q,\omega) & \hat {\bar B}^{\rm K}_{\rm v}(\bm Q,\omega) \\
0                                                         & \hat {\bar B}^{\rm A}_{\rm v}(\bm Q,\omega)
\end{array}
\right)
\nonumber\\
&=&
\left(
\begin{array}{cc}
[(\hbar\omega+i\delta)\tau_3 -\xi_{\bm Q}^{\rm ph,v}]^{-1} 
& -\pi i (1+2b_{\rm v}(\omega))\delta(\hbar\omega\tau_3-\xi_{\bm Q}^{\rm ph,v}) \\
0                  & [(\hbar\omega-i\delta)\tau_3 -\xi_{\bm Q}^{\rm ph,v}]^{-1}
\end{array}
\right),
\end{eqnarray}
\end{widetext}
is a photon Green's function in the vacuum.
The $\bm Q$-summation in the second term of Eq. (\ref{Sigph}) can be performed as, 
\begin{eqnarray}
N_{\rm t}|\Gamma_{\rm v}|^2\sum_{\bm Q}\hat {\bar B}_{\rm v}(\bm Q,\omega)
=\left(
\begin{array}{cc}
-i\kappa\tau_3 & -2i\kappa \\
0                   & i\kappa\tau_3 
\end{array}
\right).
\end{eqnarray}

From the above equations, by numerically computing $\hat{\bar\Sigma}_{\rm ph}(\bm q,\omega)$, we obtain the PL spectrum $L(\bm q,\omega)$ (Eq. (\ref{PL})), as well as the gain/absorption spectrum $S(\bm q,\omega)$ (Eq. (\ref{SW})).

As a result of our appropriate choice of diagrams, the obtained dressed Green's function $\hat {\bar D}(\bm q,\omega)$ correctly satisfies the Goldstone's theorem \cite{Goldstone1961},  (Thouless criterion \cite{Thouless1960})
\begin{eqnarray}
{\rm det}[\hat {\bar D}^{\rm R}(\bm q=0,\omega=0)]^{-1}=0,
\label{Thouless}
\end{eqnarray}
where we have used the HFB-Keldysh steady state gap equation (\ref{GAP}) in the derivation provided in Appendix \ref{App_Thouless}.
Since the pole of $\hat D^{\rm R}(\bm q,\omega)$,
\begin{eqnarray}
{\rm det}
[\hat {\bar D}^{\rm R}(\bm q,\omega_{\bm q})]^{-1}=0,
\label{mode}
\end{eqnarray}
determines the mode dispersion $\omega_{\bm q}$ \cite{Szymanska2006,Szymanska2007,Schrieffer,Hanai2017}, Eq. (\ref{Thouless}) assures the appearance of a gapless excitation (with respect to the condensate energy).

\begin{figure}
\begin{center}
\includegraphics[width=1\linewidth,keepaspectratio]{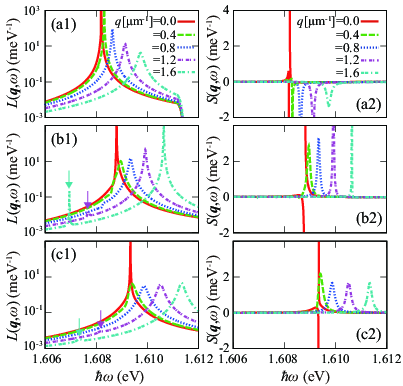}
\end{center}
\caption{(Color online) Tomographic view of (a1)--(c1) photoluminescence and (a2)--(c2) gain/absorption spectra in Fig. \ref{figPL}. [(a1) and (a2)] $\mu_{\rm b}=-4.9 {\rm meV}(\simeq \mu_{\rm b}^{c})$. [(b1) and (b2)] $\mu_{\rm b}=-3.5 {\rm meV}$. [(c1) and (c2)] $\mu_{\rm b}=-2 {\rm meV}$. The arrows points at GB peak.}
\label{fig_PL_slice}
\end{figure}

Before ending this section, let us show the steady-state properties of this system.
Figure \ref{fig_delta_mu} shows the self-consistent solution of the steady-state gap equation (\ref{GAP}) with various photon decay rate $\kappa$. 
When the pumping power $\mu_{\rm b}$ exceeds a critical value $\mu_{\rm b}^c$, a transition to a Bose-condensate phase occurs ($\Delta_0>0$), and the order parameter $\Delta_0$ increases by further increasing the pumping power.
In the equilibrium case ($\kappa=0$), a diverging behavior of $\Delta_0$ is seen at $\mu_{\rm b}=-5{\rm meV}$. 
Noting that the chemical equilibrium between the system and the bath is achieved in this limit, $\mu_{\rm b}=\mu$ (Fig. \ref{fig_delta_mu}(b)), the divergence of $\Delta_0$ is attributed to the resonant photon mediated electron-hole coupling, where the second term of Eq. (\ref{Ueff}) diverges at $\mu_{\rm b}=\mu=(\hbar\omega_{\rm cav} -E_{\rm g})/2 = -5{\rm meV}$.

When the decay rate $\kappa$ is turned on, the order parameter $\Delta_0$ is naturally suppressed by nonequilibrium effects. 
In addition to the increase of the minimal pumping power $\mu_{\rm b}^{\rm c}$ required to form a Bose-condensate, 
the divergent behavior seen in the equibrium case $\kappa=0$ is also suppressed,
due to the suppression of the resonant photon-mediated interaction.
We also see in Fig. \ref{fig_delta_mu}(b) that the ``chemical potential'' of the system $\mu$ is always smaller than the bath-chemical-potential $\mu_{\rm b}$ (i.e., $\mu_{\rm b}>\mu$) in the nonequilibrium case $\kappa>0$.
This occurs so as to induce a net electron-hole current from the bath to the system \cite{Hanai2016}, to compensate the photon loss.

\section{Photoluminescence and gain/absorption spectra of a driven-dissipative electron-hole-photon condensate}
\begin{figure}
\begin{center}
\includegraphics[width=0.6\linewidth,keepaspectratio]{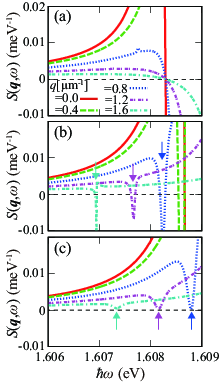}
\end{center}
\caption{(Color online) The same data of gain/absorption spectra in Figs. \ref{fig_PL_slice}(a2)--(c2), but plotted in a smaller range. (a) $\mu_{\rm b}=-4.9 {\rm meV}(\simeq \mu_{\rm b}^{c})$. (b) $\mu_{\rm b}=-3.5 {\rm meV}$. (c) $\mu_{\rm b}=-2 {\rm meV}$. The arrows point at the hole burning from GB.}
\label{fig_gain_slice_inset}
\end{figure}

\begin{figure*}
\begin{center}
\includegraphics[width=0.7\linewidth,keepaspectratio]{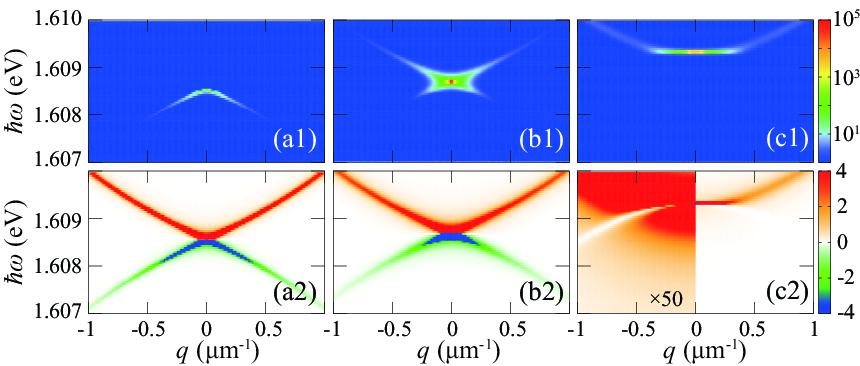}
\end{center}
\caption{(Color online) Calculated (a1)--(c1) photoluminescence and (a2)--(c2) gain/absorption spectra on resonance $\delta = 0$. (a) Equilibrium limit $\kappa=0$. (b) $\kappa=0.1~{\rm meV}$. (c) $\kappa=0.5~{\rm meV}$. 
Electron-hole density is fixed at $n_{\rm eh}=1.1\times10^3{\rm \mu m}^{-2}$.
(See Appendix \ref{App_HFB} for derivation of $n_{\rm eh}$.)
The units of the color contour are ${\rm meV}^{-1}$.
}
\label{figPL_kappa}
\end{figure*}

Our main result has already been shown in Fig. \ref{figPL}, where the bath-chemical-potential $\mu_{\rm b}$ (corresponding to the pumping power) dependence of both the PL $L(\bm q,\omega)$ and the gain/absorption spectrum $S(\bm q,\omega)$ are presented. 
The photon decay rate is set to $\kappa=0.5~{\rm meV}$, corresponding to the photon lifetime of $\tau\sim 8~{\rm ps}$.
To exhibit them in a quantitative manner, we have plotted sections through the data as a function of $\hbar\omega$ in Fig. \ref{fig_PL_slice}. 
We have also plotted a close-up view of the gain/absorption spectrum in Fig. \ref{fig_gain_slice_inset}.
As already pointed out in the introduction, the overall structure of the computed PL, including its pumping power dependence, is in agreement with experiments.
The magnitude of the condensate blue shift ($\sim{\rm meV}$), as well as the width of the momentum window of the flat spectrum ($\sim{\rm \mu m}^{-1}$), are in semiquantitative agreement with the observed PL \cite{Assmann2011,Tempel2012a}.
The visibility of the GB is strongly suppressed in both the PL and gain/absorption spectra, although a trace of it is seen as a very weak PL emission and a small but finite optical gain or a suppression of absorption band (hole burning) from the GB, as pointed by the arrows in Figs. \ref{fig_PL_slice} and \ref{fig_gain_slice_inset}, also in agreement with experiments \cite{Assmann2011,Tempel2012a,Nakayama2016,Nakayama2017,Su2017}.

\begin{figure}
\begin{center}
\includegraphics[width=1\linewidth,keepaspectratio]{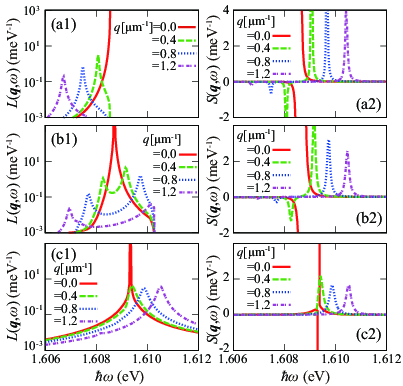}
\end{center}
\caption{(Color online) Calculated photoluminescence spectra on resonance $\delta = 0$. (a) Equilibrium limit $\kappa=0$. (b) $\kappa=0.1~{\rm meV}$. (c) $\kappa=0.5~{\rm meV}$. 
Electron-hole density is fixed at $n_{\rm eh}=1.1\times10^3{\rm \mu m}^{-2}$.
(See Appendix \ref{App_HFB} for derivation of $n_{\rm eh}$.)
}
\label{figPLkappa_slice}
\end{figure}

The suppression of the GB lies in contrast to what is obtained in the equilibrium case, for both PL and gain/absorption spectra.
In the equilibrium dilute limit (or the so-called BEC limit in the context of the BCS-BEC crossover, where the binding energy of the lower polariton is large compared to all other energy scales, i.e., $E_{\rm LP}^{\rm bind}\gg\Delta_0,\gamma$), the gain/absorption spectrum reduces to that of a repulsively interacting Bose gas within the Bogoliubov approximation \cite{Pethick,Ohashi2003}, 
\begin{eqnarray}
\bar S_{\rm eq,dil}(\bm q,\omega)
\propto
u_{\bm q}^2\delta(\hbar\omega-E^{\rm Bog}_{\bm q}) 
- v_{\bm q}^2 \delta(\hbar\omega+E^{\rm Bog}_{\bm q}),
\nonumber\\
\label{S_Bogoliubov}
\end{eqnarray}
where
\begin{eqnarray}
E_{\bm q}^{\rm Bog}=\sqrt{\varepsilon_{\bm q}^{\rm LP}(\varepsilon_{\bm q}^{\rm LP}+2U_{\rm LP}n_0)},
\end{eqnarray}
is the Bogoliubov dispersion, and 
\begin{eqnarray}
u_{\bm q}^2
&=&
\frac{1}{2}\bigg[
\frac{\varepsilon_{\bm q}^{\rm LP}+2U_{\rm LP}n_0}{E^{\rm Bog}_{\bm q}}+1
\bigg],
\\
v_{\bm q}^2
&=&
\frac{1}{2}\bigg[
\frac{\varepsilon_{\bm q}^{\rm LP}+2U_{\rm LP}n_0}{E^{\rm Bog}_{\bm q}}-1
\bigg].
\end{eqnarray}
Here, $\varepsilon_{\bm q}^{\rm LP}=\hbar^2\bm q^2 / (2m_{\rm LP})$ is the kinetic energy of the lower polariton ($m_{\rm LP}\simeq 6\times 10^{-5}m_0$ is the lower polariton mass), $n_0$ is the condensate fraction, and $U_{\rm LP}$ is the interaction between the lower polaritons. 
The positive contribution from the first term of Eq. (\ref{S_Bogoliubov}) shows that absorption occurs at the NB, while the negative contribution from the second shows that gain occurs from the GB. 
Since the ratio between $u_{\bm q}^2$ and $v_{\bm q}^2$ is order of unity in the regime $|\bm q|\lesssim \sqrt{m_{\rm LP}U_{\rm LP}n_0}/\hbar$, the intensity of the latter is comparable to the former in that regime. 
Estimating from the amount of the observed blue shift of the emission from the condensate in typical experiments \cite{Utsunomiya2008,Tempel2012a,Tempel2012b,Assmann2011,Brodbeck2016}, the magnitude of $U_{\rm LP}n_0$ should be order of a few milli-electron-volts (which is in agreement with our calculations), which gives the GB-visible region of $q\lesssim O({\rm \mu m}^{-1})$.
This is numerically demonstrated by taking the equilibrium limit of our theory in Fig. \ref{figPL_kappa}(a2) [See also the tomographic view in Fig. \ref{figPLkappa_slice}(a2)], where we obtain a strong optical gain from GB comparable to the intensity of absorption from NB, in contrast to the nonequilibrium cases seen in Figs. \ref{figPL}(a2)--(c2).

\begin{figure*}
\begin{center}
\includegraphics[width=0.7\linewidth,keepaspectratio]{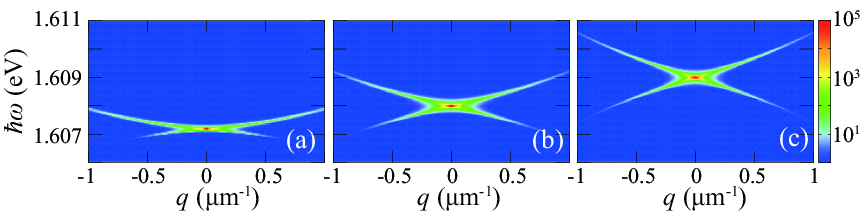}
\end{center}
\caption{(Color online) Calculated photoluminescence spectra on resonance $\delta=0$ at finite (bath) temperature $T_{\rm b}=10{\rm K}$ in the equilibrium limit $\kappa=0$, where we used the fluctuation-dissipation theorem (\ref{FDT}). (a) $\mu_{\rm b}=-8.4 {\rm meV}$. (b) $\mu_{\rm b}=-8.0 {\rm meV}$. (c) $\mu_{\rm b}=-7.5 {\rm meV}$.
The units of the color contour are ${\rm meV}^{-1}$.
}
\label{figPL_eq_T}
\end{figure*}

The GB is clearly seen in the PL in the equilibrium limit, also opposed to the nonequilibrium case. 
The PL  spectrum $L_{\rm eq,dil}(\bm q,\omega)$ in this limit can be readily calculated from the gain/absorption spectrum $S_{\rm eq,dil}(\bm q,\omega)$ by using the fluctuation-dissipation theorem \cite{Rammer},
\begin{eqnarray}
&&\bar L_{\rm eq,dil}(\bm q,\omega)
=b(\omega)\bar S_{\rm eq,dil}(\bm q,\omega)
\label{FDT}
\\
&&\propto
u_{\bm q}^2 b(\omega)\delta(\hbar\omega-E^{\rm Bog}_{\bm q})
+v_{\bm q}^2 |b(\omega)|\delta(\hbar\omega+E^{\rm Bog}_{\bm q}),
\end{eqnarray}
where $b(\omega)=[\exp{(\hbar\omega/(k_{\rm B}T_{\rm b}))}-1]^{-1}$ is the Bose distribution function.
In particular, when $T_{\rm b}=0$, since $b(\omega;T_{\rm b}=0)=-\Theta(-\omega)$ (where $\Theta(x)$ is a step function),
\begin{eqnarray}
\bar L_{\rm eq,dil}(\bm q,\omega;T_{\rm b}=0)
\propto 
v_{\bm q}^2\delta(\hbar\omega+E^{\rm Bog}_{\bm q}).
\label{PL_zerotemp}
\end{eqnarray}
Thus, when an equilibrium distribution is assumed, \textit{only} the GB is occupied in the ground state, as shown numerically in Fig. \ref{figPL_kappa}(a1) and Fig. \ref{figPLkappa_slice}(a1). This is again in stark contrast to the nonequilibrium cases [Figs. \ref{figPL}(a1)--(c1)].

We briefly note that the NB will also be occupied by thermal effects in the case of finite bath temperature. However, even in this case, the visibility of the GB is comparable to that of the NB in the small momentum regime $|\bm q|\lesssim \sqrt{m_{\rm LP}U_{\rm LP}n_0}/\hbar$.
This is demonstrated in Fig. \ref{figPL_eq_T}, where we have used the fluctuation-dissipation theorem (\ref{FDT}) to calculate PL at a realistic (bath) temperature $T_{\rm b}=10{\rm K}$. 
The strong visibility of the GB is attributed to the property that the absolute value of the Bose distribution $|b(\omega_-)|$ in the negative energy region $\omega_-<0$ is always larger than that at positive energy with the same absolute value $\omega_+=|\omega_-|$, since $|b(\omega_-)| = b(\omega_+) + 1 > b(\omega_+)$. Together with the fact that $u_{\bm q}^2/v_{\bm q}^2 = O(1)$ for $|\bm q|\lesssim \sqrt{m_{\rm LP}U_{\rm LP}n_0}/\hbar(=O({\rm \mu m^{-1}}))$, from Eq. (\ref{FDT}), the PL intensity of NB and GB are comparable.
We conclude from these considerations that the GB should have comparable visibility to the NB in the equilibrium case at small momenta.

Figures \ref{figPL_kappa}(b1), \ref{figPL_kappa}(c1), \ref{figPL_kappa}(b2), and \ref{figPL_kappa}(c2) show how the PL and gain/absorption spectra evolve as a function of the decay rate $\kappa$ [tomographic view is also shown in Figs. \ref{figPLkappa_slice}(b1), \ref{figPLkappa_slice}(c1), \ref{figPLkappa_slice}(b2), and \ref{figPLkappa_slice}(c2).].
As the decay rate $\kappa$ increases [Figs. \ref{figPL_kappa}(b1) and \ref{figPL_kappa}(c1)], the dispersion gradually evolves from a linear dispersion to a flat dispersion. 
At the same time,  the NB starts to get occupied and the emission from the GB gradually gets smaller. 
In gain/absorbtion spectra also [Figs. \ref{figPL_kappa}(a2)--(c2)], the sharp and strong optical gain from GB in the equilibrium limit gets weaker. 
These features clearly show that nonequilibrium effects strongly suppress quantum depletion.

\begin{figure*}
\begin{center}
\includegraphics[width=0.6\linewidth,keepaspectratio]{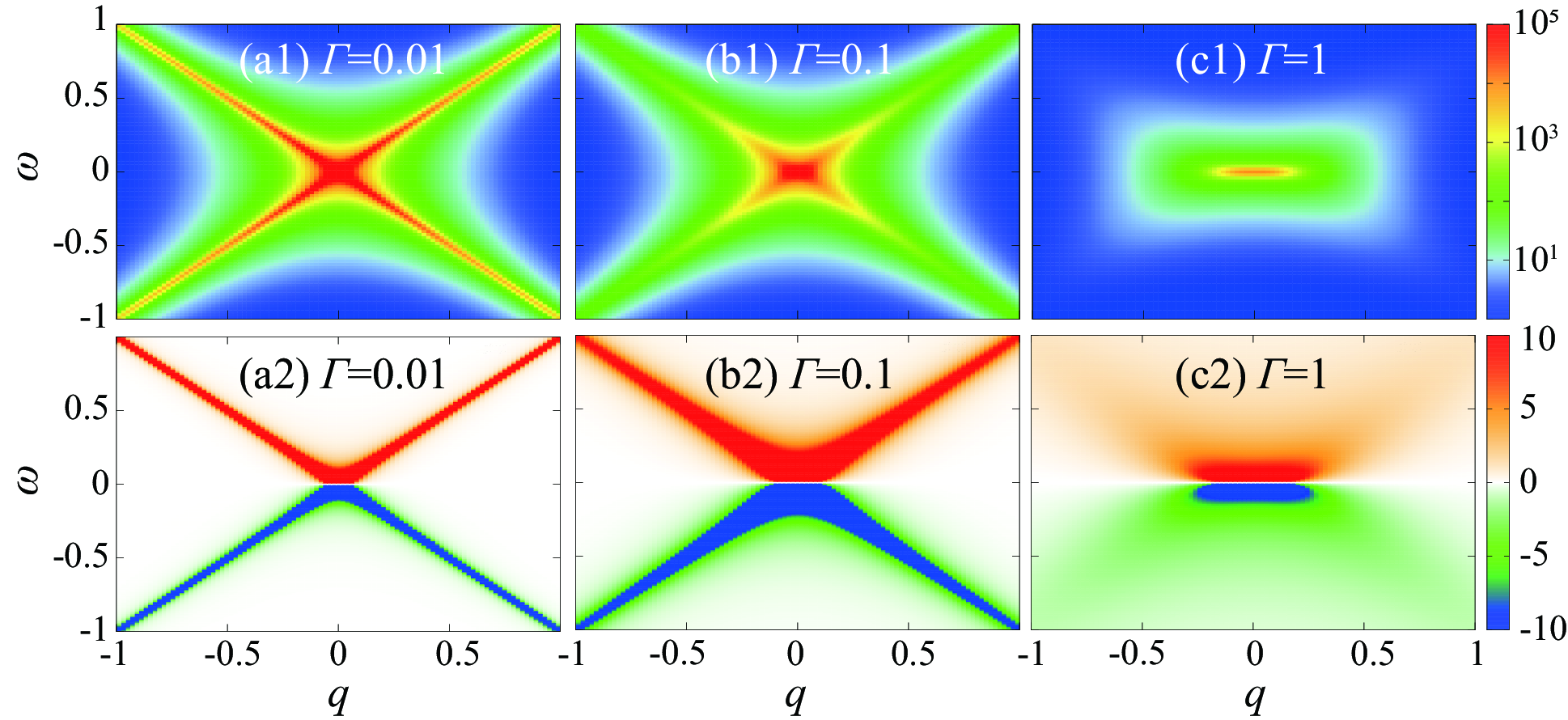}
\end{center}
\caption{(Color online) Plot of (a1)--(c1) Eq. (\ref{PL_low}) and (a2)--(c2) Eq. (\ref{gainabs_low}). [(a1) and (a2)] $\Gamma=0.01$. [(b1) and (b2)] $\Gamma=0.1$ [(c1) and (c2)] $\Gamma=1$. We set $a=c=1$.}
\label{figphotolow}
\end{figure*}

In order to understand the origin of the suppression of GB, it is useful to express the PL and gain/absorption spectra as,
\begin{widetext}
\begin{eqnarray}
\bar L(\bm q,\omega)
&=&
\frac{i}{2\pi}[\hat {\bar D}^{\rm R}(\bm q,\omega)\hat{\bar\Sigma}^{\rm <}_{\rm ph}(\bm q,\omega)\hat {\bar D}^{\rm A}(\bm q,\omega)]_{11}
\nonumber\\
&=&\frac{1}{|{\rm det}[{\bar D}^{\rm R}(\bm q,\omega)]^{-1}|^2}
\Bigg[
\left (
\begin{array}{cc}
-(\hbar\omega + i\delta)-\xi_{\bm q}^{\rm ph}-[\bar\Sigma^{\rm R}_{\rm ph}(\bm q,\omega)]_{22} & [\bar\Sigma^{\rm R}_{\rm ph}(\bm q,\omega)]_{12}\\
{}[\bar\Sigma^{\rm R}_{\rm ph}(\bm q,\omega)]_{21} & \hbar\omega + i\delta-\xi_{\bm q}^{\rm ph}-[\bar\Sigma^{\rm R}_{\rm ph}(\bm q,\omega)]_{11}
\end{array}
\right )
\nonumber\\
&\times&
\frac{i}{2\pi}\hat {\bar\Sigma}^<_{\rm ph}(\bm q,\omega)
\left (
\begin{array}{cc}
-(\hbar\omega - i\delta)-\xi_{\bm q}^{\rm ph}-[\bar\Sigma^{\rm A}_{\rm ph}(\bm q,\omega)]_{22} & [\bar\Sigma^{\rm A}_{\rm ph}(\bm q,\omega)]_{12}\\
{}[\bar\Sigma^{\rm A}_{\rm ph}(\bm q,\omega)]_{21} & \hbar\omega - i\delta-\xi_{\bm q}^{\rm ph}-[\bar\Sigma^{\rm A}_{\rm ph}(\bm q,\omega)]_{11}
\end{array}
\right )
\Bigg]_{11},
\nonumber\\
\label{PL_pole}
\\
\bar S(\bm q,\omega)
&=&
-\frac{1}{2\pi}{\rm Im}
\bigg[
\frac{1}{{\rm det}[\bar D^{\rm R}(\bm q,\omega)]^{-1}}
\Big[
-(\hbar\omega+i\delta)-\xi_{\bm q}^{\rm ph}-[\bar \Sigma_{\rm ph}^{\rm R}(\bm q,\omega)]_{22}
\Big ]
\bigg].
\label{gainabs_pole}
\end{eqnarray}
\end{widetext}
From these expressions, one can see that the pole $\hbar\omega_{\bm q}$ of Eqs. (\ref{PL_pole}) and (\ref{gainabs_pole}) determined by Eq. (\ref{mode}), which characterizes the collective motion of the condensate, directly affects PL and gain/absorption spectra.   
As discussed in Ref. \cite{Szymanska2006,Szymanska2007} within the driven-dissipative Dicke model, low energy properties of the collective mode can be studied by expanding 
${\rm det}[\hat {\bar D}^{\rm R}(\bm q,\omega)]^{-1}$ in terms of $\bm q$ and $\omega$ \cite{Hanai2017}.
By using the symmetry 
\begin{eqnarray}
{\rm det}[\hat {\bar D}^{\rm R}(\bm q,\omega)]^{-1}
&=&[{\rm det}[\hat {\bar D}^{\rm R}(-\bm q,\omega)]^{-1}],\\
{\rm det}[\hat {\bar D}^{\rm R}(\bm q,\omega)]^{-1}
&=&[{\rm det}[\hat {\bar D}^{\rm R}(\bm q,-\omega)]^{-1}]^*,
\end{eqnarray}
together with the Thouless criterion (Eq. (\ref{Thouless})), 
we can restrict the form of this expansion to (up to $O(\bm q^2,\omega^2)$),
\begin{eqnarray}
{\rm det}[\hat {\bar D}^{\rm R}(\bm q,\omega)]^{-1}
&\simeq&-a[(\hbar\omega)^2 + i\Gamma\hbar\omega  - c\bm q^2],
\label{DRanalytic}
\end{eqnarray}
where $a,c$, and $\Gamma$ are real numbers that are determined from numerical calculations.
The collective mode is then determined from Eq. (\ref{mode}) as
\begin{eqnarray}
\hbar\omega_{\bm q}\simeq -i\frac{\Gamma}{2}\pm\sqrt{c\bm q^2-\frac{\Gamma^2}{4}},
\label{diffNG}
\end{eqnarray}
which is just the diffusive Goldstone mode \cite{Szymanska2006,Szymanska2007,Wouters2007,Hanai2017}
(Eq. (\ref{diffNG_intro})).
Note that putting $\Gamma=0^+$ recovers the conventional acoustic mode realized in the equilibrium limit.

The physical picture of the diffusive Goldstone mode can be described as follows.
In the equilibrium case, the phase of a macroscopic wave function $\theta(\bm r,t)$ approximately obeys the equation of motion,
\begin{eqnarray}
\frac{\partial^2}{\partial t^2}\theta(\bm r,t)-c\nabla^2 \theta(\bm r,t)=0,
\end{eqnarray} 
giving rise to an acoustic mode with a sound velocity $\sqrt{c}$.
In the driven-dissipative case, on the other hand, particle loss from the condensate is compensated by the particle injection to the condensate.
Since the injected particles do not know the phase $\theta(\bm r,t)$ of the condensate, these pumped-in particles give rise to a nonequilibrium-induced ``friction'' for the phase,
\begin{eqnarray}
\frac{\partial^2}{\partial t^2}\theta(\bm r,t)-c\nabla^2 \theta(\bm r,t)
=-\Gamma\frac{\partial}{\partial t}\theta(\bm r,t).
\label{frictionEOM}
\end{eqnarray} 
Here, $\Gamma$ can effectively be regarded as a ``coefficient of friction'' of the condensate.
Calculating the mode dispersion of Eq. (\ref{frictionEOM}) gives the diffusive Goldstone mode (\ref{diffNG}) (apart from factor $\hbar$).

In Figs. \ref{figphotolow}(a1)--(c1), we have plotted the analytic expression derived from Eq. (\ref{DRanalytic}) of $|{\rm det} \hat {\bar D}^{\rm R}(\bm q,\omega)|^{-2}$ [denominator of Eq. (\ref{PL_pole})], 
\begin{eqnarray}
\frac{a^{-2}}{[(\hbar\omega)^2-c\bm q^2]^2+\Gamma^2{\omega}^2},
\label{PL_low}
\end{eqnarray}
and $-{\rm Im} [{\rm det} \hat {\bar D}^{\rm R}(\bm q,\omega)]^{-1}$ [denominator of Eq. (\ref{gainabs_pole})] in Figs. \ref{figphotolow}(a2)--(c2),
\begin{eqnarray}
\frac{a^{-1}\Gamma \omega}{[(\hbar\omega)^2-c\bm q^2]^2+\Gamma^2{\omega}^2},
\label{gainabs_low}
\end{eqnarray}
for various $\Gamma$ with $a=c=1$.
As one sees in the figure, Eqs. (\ref{PL_low}) and (\ref{gainabs_low}) already partially capture the decay rate $\kappa$ dependence of the calculated PL and gain/absorption spectra in Fig. \ref{figPL_kappa}, respectively.
That is, as the nonequilibrium parameter $\Gamma$ naturally increases by increase of the photon decay rate $\kappa$, a strong emission from the flat dispersion start to appear, and both the NB and the GB are strongly broadened.

The $\Gamma$ dependence shown in Fig. \ref{figphotolow} is also  similar to the pumping power dependence in Fig. \ref{figPL}. 
This is due to the property that, as the pumping power increases, the number of particles in the condensate increases. As a result, the amount of ``friction'' that the condensate suffers from gets larger to make $\Gamma$ increase, in agreement with phenomenological discussion in Ref. \cite{Wouters2007} and experiments \cite{Assmann2011,Tempel2012a,Tempel2012b,Nakayama2016,Nakayama2017,Su2017}.
We emphasize here that we have obtained these behaviors from microscopic calculations, in contrast to the phenomenological theory in Ref. \cite{Wouters2007}.

\begin{figure*}
\begin{center}
\includegraphics[width=0.66\linewidth,keepaspectratio]{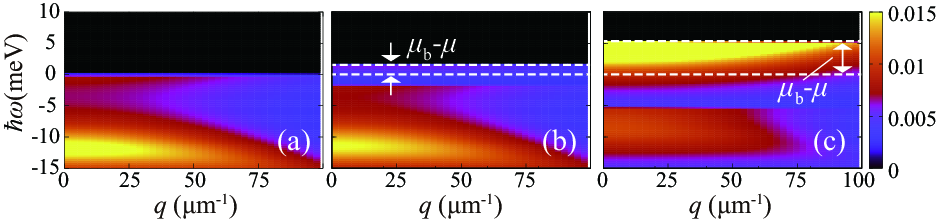}
\end{center}
\caption{(Color online) Calculated occupied spectra $L_{\rm eh}(\bm q,\omega)$ on resonance $\delta = 0$. (a) $\kappa=0$ (equilibrium limit). (b) $\kappa=0.1~{\rm meV}$. (c) $\kappa=0.5~{\rm meV}$. 
Electron-hole density is fixed at $n_{\rm eh}=1.1\times10^3{\rm \mu m}^{-2}$.
(See Appendix \ref{App_HFB} for derivation of $n_{\rm eh}$.)
The units of the color contour are ${\rm meV}^{-1}$.
}
\label{fig_spc}
\end{figure*}

\begin{figure*}
\begin{center}
\includegraphics[width=0.65\linewidth,keepaspectratio]{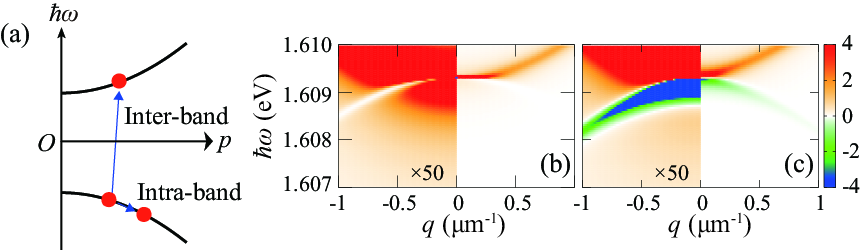}
\end{center}
\caption{(Color online) (a) Schematic explanations of inter- and intra-band excitations. [(b) and (c)] Calculated gain/absorption spectra $S(\bm q,\omega)$ on resonance $\delta = 0$ where full calculation is employed in (b), while intra-band excitations are neglected in (c). We set $\kappa=0.5{\rm meV}$ and $\mu_{\rm b}=-2{\rm meV}$
The units of the color contour in (b) and (c) are ${\rm meV}^{-1}$.
}
\label{fig_gainabs_inter}
\end{figure*}

In PL, there is also another important nonequilibrium effect, which is the redistribution of photons to higher energy by pumping and decay. 
As we have already discussed in Eq. (\ref{PL_zerotemp}), photons are occupied only at $\omega<0$ (measured from the condensate energy) in the equilibrium limit, while in nonequilibrium cases, the positive energy region $\omega>0$ start to get occupied (see e.g., Figs. \ref{figPL_kappa}(b) and \ref{figPL_kappa}(c).).
This can be explained as follows.
In the equilibrium case ($\kappa=0$) where the chemical equilibrium between the bath and the system is achieved ($\mu_{\rm b}=\mu$),
the electron and hole distribution in the bath, are given by
\begin{eqnarray}
f_{\rm b}^{\rm eq}(\omega)=\Theta(-\omega),
\end{eqnarray}
at zero bath temperature. Here, only negative energy $\omega<0$ carriers are present. 
As a result, all the electrons and holes injected to the system have negative energy. This is clearly indicated in Fig. \ref{fig_spc}(a) by plotting the occupied spectral weight function of electrons and holes, 
\begin{eqnarray}
\bar L_{\rm eh}(\bm q,\omega)
&=&\frac{1}{2\pi}\int_{-\infty}^\infty d(t-t') e^{i\omega(t-t')} 
\avg{c_{\bm p,{\rm e}}^\dagger(t')c_{\bm p,{\rm e}}(t)},
\nonumber\\
\label{Leh}
\end{eqnarray}
in the equilibrium limit, where only the lower branch $\hbar\omega_{\rm cav}=-E_{\bm p}$ (broadened by $\gamma$) is occupied.
Since a photon in the cavity is created from the pair annihilation process of an electron ($\omega_{\rm e}$) and a hole ($\omega_{\rm h}$), photons can be distributed only in the negative energy region $\omega=\omega_{\rm e}+\omega_{\rm h}<0$ where the GB lies.

In the nonequilibrium case, on the other hand,  
the bath chemical potential $\mu_{\rm b}$ gets larger than the system ``chemical potential'' $\mu$ (as shown in Fig. \ref{fig_delta_mu}(b)), as a natural consequence of having continuous injection of carriers from the bath to the system.
In this case, the bath electron and hole distribution is given by,
\begin{eqnarray}
f_{\rm b}(\omega)=\Theta(-\hbar\omega+\mu_{\rm b}-\mu),
\end{eqnarray}
where carriers with positive energy $0<\hbar\omega(<\mu_{\rm b}-\mu)$ exist.
As shown in Figs. \ref{fig_spc}(b) and (c), this results in the occupancy of the upper branch $\hbar \omega=E_{\bm p}$ (also broadened by $\gamma$), implying that pairs are partially dissociated in the nonequilibrium case \cite{Yamaguchi2012,Yamaguchi2013,Yamaguchi2015,Hanai2016,Hanai2017}.
This makes it possible for the injected electrons ($\omega_{\rm e}$) and holes ($\omega_{\rm h}$) to create photons with positive energy $0<\hbar\omega=\hbar\omega_{\rm e}+\hbar\omega_{\rm h} (< 2(\mu_{\rm b}-\mu))$, which can give a strong NB occupation that may readily exceed the GB occupation.

The strong suppression of gain from the GB is also attributed to the screening effects by the dissociated pairs.
In order to show this, following Ref. \cite{Hanai2017}, we split the lowest-order correlation function $\hat\Pi=\hat\Pi^{\rm inter}+\hat\Pi^{\rm intra}$ (Eq. (\ref{PI})) into the inter- ($\hat\Pi^{\rm inter}$) and intra-band ($\hat\Pi^{\rm intra}$) excitations, as described schematically in Fig. \ref{fig_gainabs_inter}(a).
The concrete definition is given in Appendix \ref{App_interintra}.
Here, intra-band excitations can be regarded as ``quasi-particle density fluctuations'', since they are excited by fluctuations in the particle-hole channel (where a particle and a hole are virtually created within the same branch), as in the density fluctuations in the normal state \cite{Fetter}.
Note that while the inter-band excitations can occur both in the equilibrium and nonequilibrium cases,  intra-band excitations can occur only in the latter case where the upper branch is occupied owing to the nonequilibrium-induced pair-breaking effect (Fig. \ref{fig_spc}(b)).
These nonequilibrium-induced ``quasi-particle density fluctuations'' can lead to the screening of interaction that gives rise to the optical gain from the GB.

Figures \ref{fig_gainabs_inter}(b) and \ref{fig_gainabs_inter}(c) compare the gain/absorption spectrum with and without contribution from intra-band excitations, where the latter is calculated by replacing $\hat \Pi$ to $\hat \Pi^{\rm inter}$ in the calculation of the photon self-energy Eqs. (\ref{Sigph})--(\ref{PI}).
While the fully calculated gain/absorption spectra in Fig. \ref{fig_gainabs_inter}(b) only shows hole burning from the GB, a strong optical gain in the GB is clearly present in Fig. \ref{fig_gainabs_inter}(c) when intra-band excitations are neglected.
This clearly indicates that the nonequilibrium-induced quasi-particles work as an absorption medium to screen the optical gain from GB.

\section{Visibility of ghost branch}

\begin{figure}
\begin{center}
\includegraphics[width=0.6\linewidth,keepaspectratio]{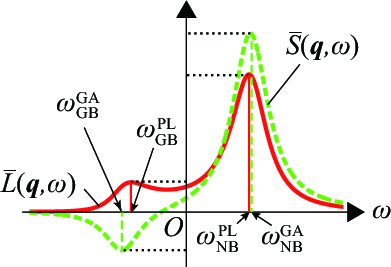}
\end{center}
\caption{(Color online) 
Definition of the NB (GB) peak position of PL $\omega_{\rm NB(GB)}^{\rm PL}(\bm q)$ and gain/absorption spectra $\omega_{\rm NB(GB)}^{\rm GA}(\bm q)$.}
\label{fig_etaPL_etaga}
\end{figure}

\begin{figure*}
\begin{center}
\includegraphics[width=0.6\linewidth,keepaspectratio]{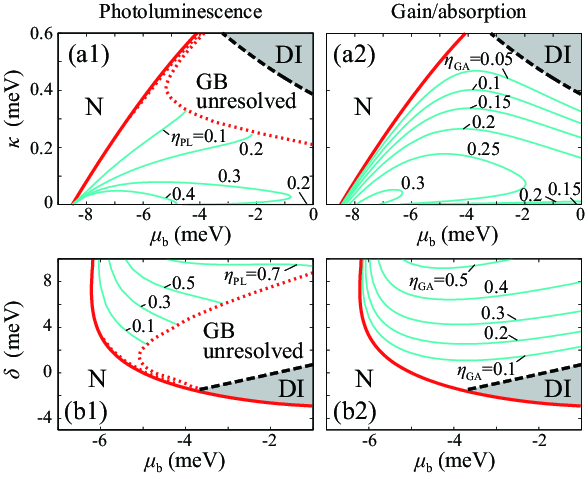}
\end{center}
\caption{(Color online) Visibility of GB in PL and gain/absorption spectra at $|\bm q|=0.4{\rm \mu m}^{-1}$. [(a1) and (b1)] $\eta_{\rm PL}(|\bm q|=0.4{\rm \mu m}^{-1})$. [(a2) and (b2)] $\eta_{\rm GA}(|\bm q|=0.4{\rm \mu m}^{-1})$.
The (red) solid line is a phase boundary between the normal phase (N) and the Bose-condensed phase. 
The (light-blue) solid thin lines are the contour of [(a1) and (b1)] $\eta_{\rm PL}(|\bm q|=0.4{\rm \mu m}^{-1})$ or [(a2) and (b2)] $\eta_{\rm GA}(|\bm q|=0.4{\rm \mu m}^{-1})$. In the ``GB unresolved'' region, GB peak in PL is absent.
In the region beyond the dashed line, denoted as ``DI'', a uniform steady-state condensate is unstable.
[(a1) and (a2)] Photon decay rate $\kappa$ and the pumping power $\mu_{\rm b}$ dependence on resonance $\delta=0$.
[(b1) and (b2)] Detuning $\delta$ and the pumping power $\mu_{\rm b}$ dependence in the case of $\kappa=0.5{\rm meV}$. 
}
\label{fig_visibility}
\end{figure*}

So far, we have shown that nonequilibrium effects strongly suppress the GB emission.
We now systematically identify the regimes where the GB becomes visible in optical quantities. 
For this purpose, we introduce the quantity
\begin{eqnarray}
\eta_{\rm PL}(\bm q)
&\equiv&
\frac{\bar L(\bm q,\omega=\omega^{\rm PL}_{\rm GB}(\bm q))}
{\bar L(\bm q,\omega=\omega^{\rm PL}_{\rm NB}(\bm q))},
\label{etaPL}
\end{eqnarray}
which takes the ratio between the PL intensity at GB $\bar L(\bm q,\omega=\omega^{\rm PL}_{\rm GB}(\bm q))$ and that at NB $\bar L(\bm q,\omega=\omega^{\rm PL}_{\rm NB}(\bm q))$, to characterize the visibility of the GB in PL.
Here, $\omega^{\rm PL}_{\rm NB(GB)}(\bm q)$ is the peak position of PL at positive (negative) energy regime $\omega>0 (\omega<0)$ for a fixed momentum $\bm q$ (Fig. \ref{fig_etaPL_etaga}).
We also define a similar quantity for gain/absorption spectra,
\begin{eqnarray}
\eta_{\rm GA}(\bm q)
&\equiv&
\frac{|\bar S(\bm q,\omega=\omega^{\rm GA}_{\rm GB}(\bm q))|}
{|\bar S(\bm q,\omega=\omega^{\rm GA}_{\rm NB}(\bm q))|}.
\label{etaGA}
\end{eqnarray}
to characterize the visibility of the gain from the GB, compared to the intensity of absorption from the NB.
Here, $\omega^{\rm GA}_{\rm NB}(\bm q)$ is the peak position of absorption from NB in the positive energy region $\omega>0$, and $\omega^{\rm GA}_{\rm GB}(\bm q)$ is that of gain (or, minimum of $S(\bm q,\omega)$) from GB in the negative energy region $\omega<0$. 
We neglect the region where gain from GB is absent (i.e., $S(\bm q,\omega=\omega_{\rm GB}^{\rm GA}(\bm q))>0$ at a given $\bm q$) in our criterion.

\begin{figure}
\begin{center}
\includegraphics[width=0.7\linewidth,keepaspectratio]{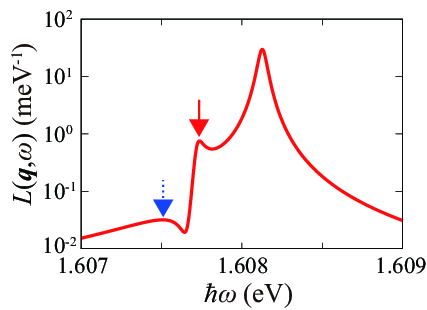}
\end{center}
\caption{(Color online) Photoluminescence at $|\bm q|=0.4{\rm \mu m}^{-1}$, where we set $\mu_{\rm b}=-6 {\rm meV}, \kappa=0.3{\rm meV}$ and $\delta=0$. The solid arrow points at the main GB peak, while the dotted arrow points at the secondary GB peak.}
\label{fig_PL_Fano_slice}
\end{figure}

Figure \ref{fig_visibility}(a1) shows $\eta_{\rm PL}(\bm q)$ at $|\bm q|=0.4{\rm \mu m}^{-1}$, as a function of the pumping power and the decay rate.
In this figure, we have plotted contours of $\eta_{\rm PL}(|\bm q|=0.4{\rm \mu m}^{-1})$, as well as the regimes where the GB peak vanishes (denoted as ``GB unresolved'' region).
We briefly note that, as shown in Fig. \ref{fig_PL_Fano_slice}, we find regimes where a secondary peak of GB appears in PL.
The appearance of the secondary peak is attributed to the occurrence of Fano resonance \cite{Fano1961}, where a resonant GB channel couples to a continuum state induced by nonequilibrium features, such as photon decay and pair-breaking effects.
In determining the ``GB unresolved'' region in Fig. \ref{fig_PL_Fano_slice}, however, we have neglected this secondary peak. 

In the equilibrium limit, as discussed earlier, only GB appears in PL, giving $\eta_{\rm PL} = \infty$ in this limit. 
Note that the equilibrium solution ($\kappa=0$) only exist at $\mu_{\rm b}=[(\hbar\omega_{\rm LP}-E_{\rm g})/2, (\hbar\omega_{\rm cav}-E_{\rm g})/2)=[-8.5{\rm meV},-5{\rm meV})$, as shown in Fig. \ref{fig_delta_mu}.
As the system gets driven away from equilibrium by the increase of the decay rate $\kappa$, $\eta_{\rm PL}(|\bm q|=0.4{\rm \mu m}^{-1})$ naturally decreases monotonically (except at small $\kappa$ with $\mu_{\rm b}\gesim 5{\rm meV}$, where no equilibrium solution is found).

\begin{figure}
\begin{center}
\includegraphics[width=0.7\linewidth,keepaspectratio]{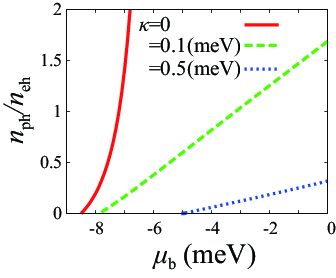}
\end{center}
\caption{(Color online) Ratio between the photon number $n_{\rm ph}$ and the electron-hole density $n_{\rm eh}$. The detuning is set to be on resonance $\delta=0$.}
\label{fig_photon}
\end{figure}

The $\mu_{\rm b}$ dependence of visibility of the GB, on the other hand, behaves non-monotonically.
That is, at relatively small decay rate ($\kappa \lesssim 0.1{\rm meV}$), $\eta_{\rm PL}(\bm q)$ possesses a maximum value at a certain pumping power $\mu_{\rm b}$. 
At larger decay rate ($\kappa \gesim 0.2{\rm meV}$), GB peak vanishes at large $\mu_{\rm b}$ (denoted as ``GB unresolved'').
These nonmonotonic behaviors can be understood as follows. 
At low pumping power close to the threshold $\mu_{\rm b}^c$, the condensate fraction increases as the pumping power increases, which gives stronger quantum depletion that makes the visibility of GB clearer.
However, at the same time, the photon number increases more rapidly than the electron and hole number, as seen in Fig. \ref{fig_photon} \cite{NOTE_photon}, due to the absence (presence) of phase filling effects of photons (electrons and holes).
This enhancement of the photon fraction has two effects; firstly, since the photon component are the ones that decay to the vacuum (while the electron and hole component thermalize the system), the nonequilibrium effects that mask the emission from GB gets more significant.
Secondly, since photons are free particles, the system gets closer to a free gas. 
Since GB emission occurs due to the repulsive interaction between polaritons, GB is suppressed as the gas becomes closer to a free photonic gas.
As a result, visibility of GB exhibits a nonmonotonic behavior as a function of $\mu_{\rm b}$.

A similar behavior is seen in gain/absorption spectra.
Figure \ref{fig_visibility}(a2) shows $\eta_{\rm GA}(|\bm q|=0.4{\rm \mu m}^{-1})$.
Apart from the difference that $\eta_{\rm GA}(|\bm q|=0.4{\rm \mu m}^{-1})$ does not diverge in the equilibrium limit, the overall behavior of $\eta_{\rm GA}(|\bm q|=0.4{\rm \mu m}^{-1})$ is similar to $\eta_{\rm PL}(|\bm q|=0.4{\rm \mu m}^{-1})$, where $\eta_{\rm GA}(|\bm q|=0.4{\rm \mu m}^{-1})$ decreases as $\kappa$ increases, and the nonmonotonical behavior as a function of the pumping power $\mu_{\rm b}$.
We note that the ``GB unresolved'' region seen in PL is absent in gain/absorption spectra, which offers an advantage to detect the GB from this quantity.

Figures \ref{fig_visibility}(b1) and (b2) show $\eta_{\rm PL}(|\bm q|=0.4{\rm \mu m}^{-1})$ and $\eta_{\rm GA}(|\bm q|=0.4{\rm \mu m}^{-1})$, respectively, as a function of the detuning $\delta$ and the pumping power $\mu_{\rm b}$.
As one sees in these figures, for both PL and gain/absorption spectra, the GB is more visible for blue detuning ($\delta>0)$.
As the system is tuned to blue (red) detuning $\delta>0 (<0)$, the energy cost to excite photons increases (decreases).
Since less (more) electron and hole injection is needed to compensate the photon loss, this drives the system to equilibrium (nonequilibrium), resulting in a stronger (weaker) emission from GB in both PL and gain/absorption spectra.

We finally note that a dynamical instability (denoted as ``DI'' in Fig. \ref{fig_visibility}) occurs in this system, in the regions where nonequilibrium effects become the most substantial (i.e., large decay rate, large pumping power, or red detuning).
Here, we have judged the stability of the steady state from the mode dispersion $\omega_{\bm q}$, determined from the mode equation (\ref{mode}).
Noting that $-{\rm Im}[\omega_{\bm q}]$ is the decay rate of fluctuations around the steady state, the steady state can be judged to be dynamically unstable when a mode exhibits negative decay rate ${\rm Im}[\omega_{\bm q}]>0$ \cite{Szymanska2006,Szymanska2007,Hanai2017}. 

This dynamical instability is triggered by an \textit{attractive} interaction between polaritons, which essentially has the same physical origin as the dynamical instability found in an electron-hole Bose condensate, discussed in our recent work \cite{Hanai2017}. 
In Ref. \cite{Hanai2017}, we have shown that the non-equilibrium induced pair-breaking effects gives rise to an anomalous virtual pair-formation processes of the broken pairs, that leads to an effective attractive channel to an exciton-exciton interaction.
As we have discussed in the previous section, a similar electron-hole pair-breaking also occurs in the present electron-hole-photon condensate.
Since the electron and hole component of polaritons is responsible for the interaction between polaritons (while the photon component is responsible for its mobility), the same scenario holds, resulting in an attractive polariton-polarition interaction which leads to a dynamical instability.

\par
\section{Summary}

To summarize, we have investigated nonequilibrium effects on optical properties of a driven-dissipative electron-hole-photon condensate.
We have formulated a combined theory of a generalized random phase approximation with the Hartree-Fock-Bogoliubov theory extended to the Keldysh formalism, that can analyze nonequilibrium effects on optical properties such as photoluminescence (PL) and gain/absorption spectra of an interacting electron-hole-photon system.
Our calculated PL is in semiquantitative agreement with experiments, where a blue shift of the condensate energy, the appearance of a diffusive Goldstone mode, and the suppression of the dispersive profile is reproduced.

We have shown that the appearance of the ghost branch (GB), which is the sign of the quantum depletion of the Bose-condensate, is strongly suppressed by nonequilibrium effects.
We have discussed that this is due to the appearance of the diffusive Goldstone mode, redistribution of photons, and screening effects by dissociatied electron-hole pairs.
It is pointed out that this suppression cannot be explained by equilibrium theories, which predicts the emission from GB to be comparable to that from the normal branch in the small momentum region, typically at $q\lesssim O({\rm \mu m}^{-1}$).
We also have shown that the GB in PL and gain/absorption spectra is more clearly seen in the blue detuning case.
The possiblity of realizing a dynamical instability, driven by dissociation of electron-hole pairs, is also pointed out.
We believe our results deepen the understanding of nonequilibrium, driven-dissipative many-body physics.

We close our paper by listing some future problems. 
Although our analysis is giving a semiquantitative agreement to experiments, there is room for improvement.
In our meanfield-based analysis for the steady state, only the zero-momentum coherent photons are concerned. 
Thus, in the normal phase (below the threshold pumping power), the calculated photon number is always zero, which is clearly not the case in experiments. 
This leads us to expect that beyond-meanfield calculations, which take into account contributions from finite momentum photons \cite{Ohashi2002,Ohashi2003,Nozieres1985,Kwong2009,Yoshioka2012} may give even better agreement to experiments in the dilute region, which remains as our future work.

In the high density region, on the other hand, the long-range nature of the realistic Coulomb interaction may play a crucial role. 
Since excitons dissociate in the region beyond the Mott density, the dynamical screening effects \cite{Zimmermann1978,Nozieres1982}, as well as pairing fluctuations \cite{Kwong2009,Yoshioka2012} may give large impact on optical properties, which is again our future work.

Lastly, the details of the physics in the dynamically unstable region (``DI'' in Fig. \ref{fig_visibility}) are unclear in the current stage of research. 
It is an interesting question to ask what would happen there after the dynamical instability take place.

\par
\begin{acknowledgements}
We thank A. Edelman, M. Yamaguchi, K. Kamide, and T. Ogawa for useful discussions. This work was supported by the KiPAS project at Keio University. 
R.H. was supported by Grant-in-Aid for JSPS fellows (Grant No. 15J02513). 
Y.O. was supported by Grants-in-Aid for Scientific Research from MEXT and JSPS in Japan (Grants No. JP15K00178, No. JP15H00840, and No. JP16K05503). 
Work at Argonne National Laboratory is supported by the Materials Sciences and Engineering Division, Basic Energy Sciences, Office of Science, US DOE under Contract No. DE-AC02-06CH11357.
\end{acknowledgements}
\begin{appendix}

\section{Choice of coupling constants $U$ and $g$}\label{App_parameter}

In this paper, we choose the magnitude of coupling constants $U$ and $g$ that reproduces the exciton and polariton binding energy in a GaAs quantum well structure embedded to a microcavity.
We first determine the magnitude of $U$, by considering the case $g=0$ (corresponding to an electron-hole gas without the microcavity structure) in the dilute equilibrium limit ($\Delta_0=\kappa=0$). 
In this case, the system is well described by a free exciton gas, where the chemical potential of electrons or holes (which is in the chemical equilibrium with the bath $\mu_{\rm b}=\mu$) is known to be negative and have an absolute value of half the binding energy, i.e., $\mu=-E^{\rm bind}_{\rm X}/2 = -5{\rm meV}$. 
This physically means that, when an electron and a hole is added to the system, an exciton is formed to earn its binding energy.
We tune the coupling constant $U$ to be consistent with this picture by solving the gap equation (\ref{GAP}) in this limit, (Note that $F_+(\omega)=0$ when $\mu_{\rm b}=\mu$.) 
\begin{widetext}
\begin{eqnarray}
\frac{1}{U}= \sum_{\bm p}\int \frac{\hbar d\omega} {\pi}
\frac{F_-(\omega;T_{\rm b}=0) \hbar\omega}
{[(\hbar\omega - (\varepsilon_{\bm p}^{\rm eh}+E^{\rm bind}_{\rm X}/2))^2 + \gamma^2]
[(\hbar\omega + (\varepsilon_{\bm p}^{\rm eh}+E^{\rm bind}_{\rm X}/2))^2 + \gamma^2]
},
\label{excitonbind}
\end{eqnarray}
where we find $U=5.2{\rm meV/\mu m}^2$ for a cutoff momentum $k_c=2\pi /a = 1360{\rm \mu m}^{-1}$.

\begin{figure}
\begin{center}
\includegraphics[width=0.37\linewidth,keepaspectratio]{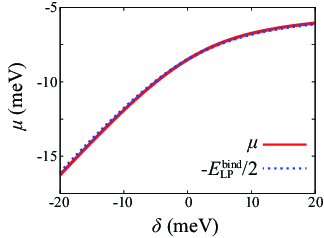}
\end{center}
\caption{(Color online) The chemical potential $\mu$ in the dilute equilibrium limit ($\Delta_0=\kappa=0$). The dotted line shows $-E^{\rm bind}_{\rm LP}(\delta)/2$, where $E^{\rm bind}_{\rm LP}(\delta)$ is given by Eq. (\ref{ELP}).}
\label{fig_ELP}
\end{figure}

Similarly, the dipole coupling $g$ is determined by using the property that an electron-hole-photon gas in the dilute equilibrium limit is described by a free lower polariton gas \cite{Deng2010}.
We demand the chemical potential $\mu(=\mu_{\rm b})$ to satisfy $\mu=-E^{\rm bind}_{\rm LP}(\delta)/2=-8.5{\rm meV}$ on resonance $\delta=0$, by solving the gap equation,
\begin{eqnarray}
&&\frac{1}{U_{\rm eff}(\delta=0,\kappa=0)} 
= \sum_{\bm p}\int \frac{\hbar d\omega} {\pi}
\frac{F_-(\omega) \hbar\omega}
{[(\hbar\omega - (\varepsilon_{\bm p}^{\rm eh}+E^{\rm bind}_{\rm LP}(\delta=0)/2))^2 + \gamma^2]
[(\hbar\omega + (\varepsilon_{\bm p}^{\rm eh}+E^{\rm bind}_{\rm LP}(\delta=0)/2))^2 + \gamma^2]
},
\label{LPbind}
\end{eqnarray}
where
\begin{eqnarray}
E^{\rm bind}_{\rm LP}(\delta)=
\frac{1}{2}
\Big[
-\delta + 2E^{\rm bind}_{\rm X}+\sqrt{\delta^2+4g_{\rm R}^2}
\Big],
\label{ELP}
\end{eqnarray}
is the binding energy of the lower polariton \cite{Deng2010}.
From this equation, we find $g=1.7{\rm meV/\mu m}^2$.
Although Eq. (\ref{LPbind}) only deals with the on resonance case ($\delta=0$), 
as shown in Fig. \ref{fig_ELP}, we have checked that our choice of $U$ and $g$ approximately satisfy $\mu=-E^{\rm bind}_{\rm LP}(\delta)/2$ in the dilute equilibrium limit in a wide range of detuning parameter $\delta$.


\section{Hartree-Fock-Bogoliubov-Keldysh theory of a driven-dissipative electron-hole-photon condensate}\label{App_HFB}

Here, we present the Hartree-Fock-Bogoliubov-Keldysh theory of a driven-dissipative electron-hole-photon condensate \cite{Yamaguchi2012,Yamaguchi2013,Yamaguchi2015}.
The central quantity in this formalism is the Nambu-Keldysh single-particle Green's function of electrons and holes, defined by, 
\begin{eqnarray}
\hat G(\bm p,t-t')
&=&
\left(
\begin{array}{cc}
\hat G_{aa}(\bm p,t-t') & \hat G_{ab}(\bm p,t-t') \\
\hat G_{ba}(\bm p,t-t') & \hat G_{bb}(\bm p,t-t')
\end{array}
\right)=
\left(
\begin{array}{cc}
\hat G^{\rm R}(\bm p,t-t') & \hat G^{\rm K}(\bm p,t-t') \\
0                                 & \hat G^{\rm A}(\bm p,t-t')
\end{array}
\right)
\nonumber\\
&=&
-i\left(
\begin{array}{cc}
\theta(t-t')\avg{\{\hat \Psi_{\bm p}(t)\diamondcomma \hat \Psi^\dagger_{\bm p}(t')\}} 
& \avg{\hat \Psi_{\bm p}(t)\diamond \hat \Psi^\dagger_{\bm q}(t') - \hat \Psi^\dagger_{\bm p}(t')\diamond \hat \Psi_{\bm p}(t)}\\
0 
& \theta(t'-t)\avg{\{\hat \Psi_{\bm p}(t)\diamondcomma \hat \Psi^\dagger_{\bm p}(t')\}}
\end{array}
\right).
\label{G_eh}
\end{eqnarray}
which obeys the Dyson's equation \cite{Rammer}, 
\begin{eqnarray}
\hat {G}_{\alpha,\beta}(\bm p,\omega)
=\hat {G}^0_{\alpha,\beta}(\bm p,\omega)+\hat {G}^0_{\alpha,\alpha'}(\bm p,\omega)\hat {\Sigma}_{\alpha',\beta'}(\bm p,\omega)\hat {G}_{\beta',\beta}(\bm p,\omega).
\label{Dyson}
\end{eqnarray}
Here, we have introduced a Nambu representation of the electron-hole operator, 
\begin{eqnarray}
\hat\Psi_{\bm p}
=\left(
\begin{array}{c}
c_{\bm p,{\rm e}}\\
c^\dagger_{-\bm p,{\rm h}}
\end{array}
\right)
\equiv
\left(
\begin{array}{c}
\Psi_{\bm p,1}\\
\Psi_{\bm p,2}
\end{array}
\right),
\end{eqnarray}
as well as the operations \cite{Rammer},
\begin{eqnarray}
\big(
\hat \Psi_{\bm p}(t)\diamond \hat \Psi_{\bm p}^\dagger(t')
\big)_{s,s'}
&\equiv&
\Psi_{\bm p,s}(t)\Psi_{\bm p,s'}^\dagger(t')
=
\left(
\begin{array}{cc}
c_{{\bm p},{\rm e}}(t)   c_{{\bm p},{\rm e}}^\dagger(t')  & c_{\bm p,{\rm e}}(t)c_{-\bm p,{\rm h}}(t')\\
c_{-\bm p,{\rm h}}^\dagger(t) c_{\bm p,{\rm e}}^\dagger(t') & c_{-\bm p,{\rm h}}^\dagger(t) c_{-\bm p,{\rm h}}(t')
\end{array}
\right)_{s,s'}
,\\
\big(
\hat \Psi^\dagger_{\bm p}(t')\diamond \hat \Psi_{\bm p}(t)
\big)_{s,s'}
&\equiv&
\Psi^\dagger_{\bm p,s'}(t')\Psi_{\bm p,s}(t)
=\left(
\begin{array}{cc}
c^\dagger_{\bm p,{\rm e}}(t') c_{\bm p,{\rm e}}(t)              & c_{-\bm p,{\rm h}}(t') c_{\bm p,{\rm e}}(t)\\
c_{\bm p,{\rm e}}^\dagger(t') c_{-\bm p,{\rm h}}^\dagger(t) & c_{-\bm p,{\rm h}}(t') c_{-\bm p,{\rm h}}^\dagger(t) 
\end{array}
\right)_{s,s'}.
\end{eqnarray}
We have assumed that the system is in a uniform steady state.
Below, we employ the gauge transformation described in the paragraph below Eq. (\ref{Ep}), in order to formally eliminate the time dependence of the order parameter in Eq. (\ref{Delta}). 
A free single-particle Green's function is given by
\begin{eqnarray}
\hat {\bar G}^0(\bm p,\omega)
&=&
\left(
\begin{array}{cc}
\hat {\bar G}^0_{aa}(\bm p,\omega) & \hat {\bar G}^0_{ab}(\bm p,\omega) \\
\hat {\bar G}^0_{ba}(\bm p,\omega) & \hat {\bar G}^0_{bb}(\bm p,\omega)
\end{array}
\right)=
\left(
\begin{array}{cc}
\hat {\bar G}^{0{\rm R}}(\bm p,\omega) & \hat {\bar G}^{0{\rm K}}(\bm p,\omega) \\
0                                         & \hat {\bar G}^{0{\rm A}}(\bm p,\omega)
\end{array}
\right)
\nonumber\\
&=&
\left(
\begin{array}{cc}
[\hbar\omega+i\delta -\xi_{\bm p}\tau_3]^{-1} & -2\pi i \tau_3 (1-2f(\hbar\omega\tau_3)) \delta(\hbar\omega-\xi_{\bm p}\tau_3) \\
0                  & [\hbar\omega-i\delta -\xi_{\bm p}\tau_3]^{-1}
\end{array}
\right),
\end{eqnarray}
and
\begin{eqnarray}
\hat {\bar \Sigma}(\bm p,\omega)
&=&
\left(
\begin{array}{cc}
\hat {\bar \Sigma}_{aa}(\bm p,\omega) & \hat {\bar \Sigma}_{ab}(\bm p,\omega) \\
\hat {\bar \Sigma}_{ba}(\bm p,\omega) & \hat {\bar \Sigma}_{bb}(\bm p,\omega)
\end{array}
\right)=
\left(
\begin{array}{cc}
\hat {\bar \Sigma}^{\rm R}(\bm p,\omega) & \hat {\bar \Sigma}^{\rm K}(\bm p,\omega) \\
0                                                      & \hat{\bar \Sigma}^{\rm A}(\bm p,\omega)
\end{array}
\right),
\end{eqnarray}
is the self-energy that incorporates interaction effects in a nonequilibrium situation.

\begin{figure}
\begin{center}
\includegraphics[width=0.45\linewidth,keepaspectratio]{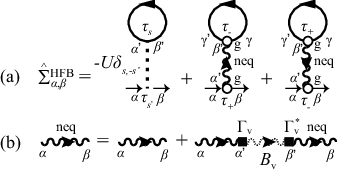}
\end{center}
\caption{(Color online) Diagramatic representation of (a) HFB self-energy $\hat\Sigma^{\rm HFB}$ and (b) photon Green's function coupled to a vacuum $D_{\rm neq}^0$.
Here, the solid line describes the single-particle Green's function $\hat G$, and the wavy line denoted with (without) ``neq'' describes $D_{\rm neq}^0 ~(D^0)$. The dotted line and the open circle represent the electron-hole coupling $-U$ and photon-electron-hole dipole coupling $g$, respectively. The dashed-wavy line represent the vacuum photon Green's function $B_{\rm v}$. The solid rectangle describes the tunneling $\Gamma_{\rm v}$.}
\label{figdiagram_HFB}
\end{figure}

The self-energy in this HFB-Keldysh framework is given by $\hat\Sigma=\hat\Sigma^{\rm HFB}+\hat\Sigma^{\rm env}$. 
The HFB self-energy $\hat\Sigma^{\rm HFB}$, represented diagramatically in Fig. \ref{figdiagram_HFB}(a), is given by \cite{Schrieffer,Ohashi2003,Yamaguchi2012,Yamaguchi2013,Yamaguchi2015}
\begin{eqnarray}
\hat{\bar \Sigma}_{\alpha,\beta}^{\rm HFB}(\bm p, \omega) 
&=&iU\sum_{\bm p'}\int_{-\infty}^\infty \frac{\hbar d\omega}{2\pi}
\sum_{\alpha',\beta'=\pm}\sum_{s,s'=\pm}
\eta^{\alpha,\alpha'}_{\beta,\beta'}\delta_{s,-s'}
{\rm Tr}[\tau_s \hat {\bar G}_{\beta',\alpha'}(\bm p',\omega)]\tau_{s'}
\nonumber\\
&-&ig^2
\sum_{\bm p'}\int_{-\infty}^\infty\frac{\hbar d\omega}{2\pi}
\sum_{\alpha',\beta'=\pm}\sum_{\gamma,\gamma'=\pm}
\gamma_{\alpha,\beta}^{\alpha'} [{\bar D}_{\rm neq}^0]_{\alpha',\beta'}(\bm 0,0)\tilde\gamma_{\gamma',\gamma}^{\beta'}
{\rm Tr}[\tau_- \hat {\bar G}_{\gamma,\gamma'}(\bm p',\omega)]\tau_{+} 
\nonumber\\
&-&ig^2
\sum_{\bm p'}\int_{-\infty}^\infty\frac{\hbar d\omega}{2\pi}
\sum_{\alpha',\beta'=\pm}\sum_{\gamma,\gamma'=\pm}
\tilde\gamma_{\alpha,\beta}^{\alpha'} [\bar D_{\rm neq}^0]_{\beta',\alpha'}(\bm 0,0)\gamma_{\gamma',\gamma}^{\beta'}
{\rm Tr}[\tau_+ \hat {\bar G}_{\gamma,\gamma'}(\bm p',\omega)]\tau_{-} 
\nonumber\\
&=&
iU\sum_{\bm p'}\int_{-\infty}^\infty \frac{\hbar d\omega}{2\pi}\sum_{s,s'=\pm}\frac{1}{2}\delta_{s,-s'}
\left(
\begin{array}{cc}
{\rm Tr}[\tau_s \hat {\bar G}^{\rm K}(\bm p',\omega)] & 0 \\
0 & {\rm Tr}[\tau_s \hat {\bar G}^{\rm K}(\bm p',\omega)]
\end{array}
\right)_{\alpha,\beta}
\tau_{s'}
\nonumber\\
&-&ig^2
\sum_{\bm p'}\int_{-\infty}^\infty \frac{\hbar d\omega}{2\pi}
\frac{1}{2}
{\bar D}_{\rm neq}^{0{\rm R}}(\bm 0,0)
\left(
\begin{array}{cc}
{\rm Tr}[\tau_- \hat {\bar G}^{\rm K}(\bm p',\omega)] & 0 \\
0 & {\rm Tr}[\tau_- \hat {\bar G}^{\rm K}(\bm p',\omega)]
\end{array}
\right)_{\alpha,\beta}
\tau_{+}
\nonumber\\
&-&ig^2
\sum_{\bm p'}\int_{-\infty}^\infty \frac{\hbar d\omega}{2\pi}
\frac{1}{2}
{\bar D}_{\rm neq}^{0{\rm A}}(\bm 0,0)
\left(
\begin{array}{cc}
{\rm Tr}[\tau_+ \hat {\bar G}^{\rm K}(\bm p',\omega)] & 0 \\
0 & {\rm Tr}[\tau_+ \hat {\bar G}^{\rm K}(\bm p',\omega)]
\end{array}
\right)_{\alpha,\beta}
\tau_{-}.
\label{SigHFB}
\end{eqnarray}
\end{widetext}

The first term in Eq. (\ref{SigHFB}) describes the direct electron-hole interaction effects  $-U$, while the second and the third describe effects by an effective interaction that arises from the second-order processes of emission and absorption of photons.
In the latter, the decay process of photons is incorporated in $D_{\rm neq}^{0{\rm R/A}}(\bm q,\omega)$, diagrammatically represented in Fig. \ref{figdiagram_HFB}(b), given by 
\begin{eqnarray}
{\bar D}_{\rm neq}^{0{\rm R}}(\bm q,\omega)
&=&
{\bar D}^{0{\rm R}}(\bm q,\omega)+{\bar D}^{0{\rm R}}(\bm q,\omega)
\bar\Sigma_{\rm phv}^{\rm R}(\omega)
{\bar D}_{\rm neq}^{0{\rm R}}(\bm q,\omega),
\nonumber\\
\label{DR0tilde}
\\
{\bar D}_{\rm neq}^{0{\rm A}}(\bm q,\omega)
&=&
[{\bar D}_{\rm neq}^{0{\rm R}}(\bm q,\omega)]^\dagger.
\label{DA0tilde}
\end{eqnarray}
Here,
\begin{eqnarray}
{\bar \Sigma}_{\rm phv}^{\rm R}(\omega)
&=&
N_{\rm t}|\Gamma_{\rm v}|^2\sum_{\bm Q}{\bar B}_{\rm v}^{\rm R}(\bm Q,\omega)=- i \kappa,
\label{Sigrphv}
\\
{\bar \Sigma}_{\rm phv}^{\rm A}(\omega)
&=&
[{\bar \Sigma}_{\rm phv}^{\rm R}(\omega)]^\dagger
=i\kappa,
\label{Sigaphv}
\end{eqnarray}
describes the decay of cavity photons by tunneling to the vacuum within the second-order Born approximation \cite{note_Born}, where
\begin{eqnarray}
{\bar B}_{\rm v}^{\rm R}(\bm Q,\omega)
&=&
[\hat {\bar B}^{\rm v}_{11}(\bm Q,\omega)]^{\rm R}
=
\frac{1}{\hbar\omega-\xi_{\bm Q}^{\rm ph,v}+ i\delta},
\\
{\bar B}_{\rm v}^{\rm A}(\bm Q,\omega)
&=&
[{\bar B}_{\rm v}^{\rm R}(\bm Q,\omega)]^\dagger,
\end{eqnarray}
is the vacuum photon propagator. 
We have taken the random average over the tunneling points $\bm r_i$ and $\bm R_i$ in obtaining the first equality of Eq. (\ref{Sigrphv}). 
In the second equality, we have assumed a white vacuum with a constant density of states (See Eq. (\ref{kappa}) for the definition of $\kappa$).
As is clear from from Eq. (\ref{Sigrphv}), the coupling to the vacuum induce the photon lifetime of 
\begin{eqnarray}
\tau=\frac{2\pi\hbar}{\kappa}.
\end{eqnarray}
Thus, $\kappa$ can be interpreted as the decay rate of photons from the cavity.
${\bar D}^0$ is a free Green's function of cavity photons, 
\begin{widetext}
\begin{eqnarray}
{\bar D}^0(\bm q,\omega)
=
\hat {\bar D}^0_{11}(\bm q,\omega)
&=&
\left(
\begin{array}{cc}
{\bar D}^0_{aa}(\bm q,\omega) & {\bar D}^0_{ab}(\bm q,\omega) \\
{\bar D}^0_{ba}(\bm q,\omega) & {\bar D}^0_{bb}(\bm q,\omega)
\end{array}
\right)=
\left(
\begin{array}{cc}
{\bar D}^{0{\rm R}}(\bm q,\omega) & {\bar D}^{0{\rm K}}(\bm q,\omega) \\
0                                           & {\bar D}^{0{\rm A}}(\bm q,\omega)
\end{array}
\right)
\nonumber\\
&=&
\left(
\begin{array}{cc}
[\hbar\omega+i\delta -\xi_{\bm q}^{\rm ph}]^{-1} & -\pi i (1+2b_{\rm ph}(\omega))\delta(\hbar\omega-\xi_{\bm q}^{\rm ph}) \\
0                  & [\hbar\omega-i\delta -\xi_{\bm q}^{\rm ph}]^{-1}
\end{array}
\right).
\label{D0}
\end{eqnarray}
From Eqs. (\ref{DR0tilde})-(\ref{D0}), we obtain
\begin{eqnarray}
{\bar D}_{\rm neq}^{0\rm {R}}(\bm q,\omega) 
&=&
\frac{1}{\hbar\omega-\xi_{\bm q}^{\rm ph} + i\kappa},
\label{D0R}
\\
{\bar D}_{\rm neq}^{0\rm {A}}(\bm q,\omega) 
&=&
[{\bar D}_{\rm neq}^{0\rm {R}}(\bm q,\omega) ]^\dagger.
\end{eqnarray}

The HFB self-energy (Eq. (\ref{SigHFB})) is thus obtained as 
\begin{eqnarray}
\hat{\bar{\Sigma}}_{\alpha,\beta}^{\rm HFB}(\bm p, \omega) 
&=&
i
\sum_{\bm p'}\int_{-\infty}^\infty \frac{\hbar d\omega}{2\pi}
\frac{1}{2}
\left(
\begin{array}{cc}
{\rm Tr}[\tau_- U_{\rm eff}\hat {\bar G}^{\rm K}(\bm p',\omega)] & 0 \\
0 & {\rm Tr}[\tau_- U_{\rm eff}^* \hat {\bar G}^{\rm K}(\bm p',\omega)]
\end{array}
\right)_{\alpha,\beta}
\tau_{+}
\nonumber\\
&-&i
\sum_{\bm p'}\int_{-\infty}^\infty \frac{\hbar d\omega}{2\pi}
\frac{1}{2}
\left(
\begin{array}{cc}
{\rm Tr}[\tau_+ U_{\rm eff}^* \hat {\bar G}^{\rm K}(\bm p',\omega)] & 0 \\
0 & {\rm Tr}[\tau_+ U_{\rm eff}  \hat {\bar G}^{\rm K}(\bm p',\omega)]
\end{array}
\right)_{\alpha,\beta}
\tau_{-},
\label{SigHFB_Ueff}
\end{eqnarray}
\end{widetext}
where an effective interaction $U_{\rm eff}$ is given by Eq. (\ref{Ueff}).
The retarded component of HFB self-energy (Eq. (\ref{SigHFB_Ueff})) can also be written as
\begin{eqnarray}
\big[\hat{\bar \Sigma}^{\rm HFB}(\bm p, \omega)\big]^{\rm R}
&=&-U_{\rm eff}
\sum_{\bm p'}\avg{\bar c_{-\bm p',{\rm h}}\bar c_{\bm p',{\rm e}}}\tau_+ 
\nonumber\\
&-&
U_{\rm eff}^*\sum_{\bm p'}\avg{\bar c^\dagger_{\bm p',{\rm e}}\bar c^\dagger_{-\bm p',{\rm h}}}\tau_-,
\label{SigHFBDel_B}
\end{eqnarray}
where we have used the relation,
\begin{eqnarray}
\avg{\bar c_{-\bm p,{\rm h}} \bar c_{\bm p,{\rm e}}}
&=&
 \frac{-i}{2} i
[\avg{\bar c_{-\bm p,{\rm h}} \bar c_{\bm p,{\rm e}}}
-\avg{\bar c_{\bm p,{\rm e}} \bar c_{-\bm p,{\rm h}}}]
\nonumber\\
&=&
-\frac{i}{2}\int_{-\infty}^\infty \frac{d\omega}{2\pi}{\bar G}^{\rm K}_{12}(\bm p,\omega).
\end{eqnarray}

\begin{figure}
\begin{center}
\includegraphics[width=0.6\linewidth,keepaspectratio]{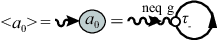}
\end{center}
\caption{(Color online) Diagramatic representation of photon amplitude $\avg{a_0}$.
Here, the solid line describes $\hat G$, and the wavy line denoted with ``neq'' describes $D_{\rm neq}^0$. The open circle represent the photon-electron-hole dipole coupling $g$.}
\label{figdiagram_a0}
\end{figure}
As diagramatically shown in Fig. \ref{figdiagram_a0}, we can relate $\avg{\bar a_0}$ and $\sum_{\bm p'}\avg{\bar c_{-\bm p',{\rm h}} \bar c_{\bm p',{\rm e}}}$ as, 
\begin{eqnarray}
&&\avg{{\bar a_0}} 
=-ig{\bar D}_{\rm neq}^{0{\rm R}}(\bm 0,0)
\sum_{\bm p'}\int_{-\infty}^\infty \frac{d\omega}{2\pi}
{\bar G}^{\rm K}_{12}(\bm p',\omega)
\nonumber\\
&&= -\frac{g}{\hbar\omega_{\rm cav} - 2\mu - E_{\rm g} - i\kappa}
\sum_{\bm p'}\avg{\bar c_{-\bm p',{\rm h}} \bar c_{\bm p',{\rm e}}},
\label{Langevin}
\end{eqnarray}
by applying the Wick's theorem. 
This simplifies Eq. (\ref{SigHFBDel_B}) to 
\begin{eqnarray}
\big[\hat{\bar \Sigma}^{\rm HFB}(\bm p, \omega)\big]^{\rm R}
&=&
-[U\sum_{\bm p'}\avg{\bar c_{-\bm p',{\rm h}}\bar c_{\bm p',{\rm e}}}-g\avg{\bar{a_0}}]\tau_+
\nonumber\\
&-&
[U\sum_{\bm p'}\avg{\bar c^\dagger_{\bm p',{\rm e}}\bar c^\dagger_{-\bm p',{\rm h}}}-g\avg{\bar {a_0}}^*]\tau_-
\nonumber\\
&=&-\Delta_0\tau_1.
\label{SigHFBDel}
\end{eqnarray}

\begin{figure}
\begin{center}
\includegraphics[width=0.5\linewidth,keepaspectratio]{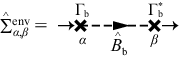}
\end{center}
\caption{(Color online) Diagramatic representation of the self-energy of the bath-system coupling $\hat\Sigma^{\rm env}$. 
The dashed line represent the bath Green's function $\hat B_{\rm b}$, and the cross describes the tunneling $\Gamma_{\rm b}$.}
\label{figdiagram_env}
\end{figure}

The pumping of electrons and holes from the bath compensates the photon decay. 
Figure \ref{figdiagram_env} gives the diagram of the self-energy that describes these processes,
\begin{eqnarray}
&&\hat{\bar \Sigma}^{\rm env}_{\alpha,\beta}({\bm p},\omega) 
=
N_{\rm t}|\Gamma_{\rm b}|^2
\sum_{\bm P}\hat {\bar B}^{\rm b}_{\alpha,\beta}({\bm P},\omega)
\nonumber\\
&&=
\left(
\begin{array}{cc}
-i\gamma & 
-2i\tau_3\gamma [1-2f_{\rm b}(\omega\tau_3)]
\\
0 & i\gamma
\end{array}
\right)_{\alpha,\beta},
\label{Sigenv}
\end{eqnarray}
within the second-order Born approximation \cite{note_Born}.
Again, we have taken the random average over the tunneling points $\bm r_i$ and $\bm R_i$, and assumed a white bath with a constant density of states $\rho_{\rm b}$.
As $\gamma$ gives the decay rate of a quasi-particle induced by coupling to the bath, this quantity can be interpreted as the thermalization rate.

The retarded component of the Green's function is obtained from the Dyson's equation (\ref{Dyson}) and the retarded component of the self-energies (\ref{SigHFBDel}), (\ref{Sigenv}), as
\begin{eqnarray}
\hat {\bar G}^{\rm R}(\bm p,\omega)
&=&[(\hbar\omega + i\gamma) \bm 1 - \xi_{\bm p}\tau_3+\Delta_0 \tau_1]^{-1}
\nonumber\\
&=&
\frac{(\hbar\omega + i\gamma)\bm 1+\xi_{\bm p}\tau_3-\Delta_0\tau_1}
{(\hbar\omega + i\gamma)^2-E_{\bm p}^2},
\label{Gr}
\\
\hat {\bar G}^{\rm A}(\bm p,\omega)
&=&[\hat {\bar G}^{\rm R}(\bm p,\omega)]^\dagger.
\end{eqnarray}
The Dyson's equation (\ref{Dyson}) also gives the Keldysh component of the Green's function as
\begin{eqnarray}
\hat {\bar G}^{\rm K}(\bm p,\omega)
&=&
\hat {\bar G}^{\rm R}(\bm p,\omega)\hat {\bar \Sigma}^{\rm K}(\omega)\hat {\bar G}^{\rm A}(\bm p,\omega)
\nonumber\\
&+&
[1+\hat{\bar G}^{\rm R}(\bm p,\omega)\hat{\bar\Sigma}^{\rm R}(\bm p,\omega)]
G^{0{\rm K}}(\bm p,\omega)
\nonumber\\
&\times&
[1+\hat{\bar\Sigma}^{\rm A}(\bm p,\omega)\hat{\bar G}^{\rm A}(\bm p,\omega)],
\label{Gk}
\end{eqnarray}
where the second term of Eq. (\ref{Gk}) can be shown to vanish.
Using Eq. (\ref{Gk}), a self-consistent condition between Eqs. (\ref{SigHFB_Ueff}) and (\ref{SigHFBDel}) can be obtained as \cite{Szymanska2006,Szymanska2007,Yamaguchi2012,Yamaguchi2013,Yamaguchi2015,Hanai2016,Hanai2017},   
\begin{eqnarray}
\Delta_0 
&= &
\Big[
U+\frac{g}{\hbar\omega_{\rm cav} - 2\mu - E_{\rm g} - i\kappa}
\Big]
\sum_{\bm p}\avg{\bar c_{-\bm p,{\rm h}}\bar c_{\bm p,{\rm e}}}
\nonumber\\
&=&
-iU_{\rm eff}\sum_{\bm p}\int_{-\infty}^\infty \frac{d\omega}{2\pi}
\frac{1}{2}G_{12}^{\rm K}(\bm p,\omega),
\label{GAP_Gk}
\end{eqnarray}
which gives the nonequilibrium steady-state gap equation (\ref{GAP}).

The occupied spectral weight function of electrons and holes $\bar L_{\rm eh}(\bm q,\omega)$, defined in Eq. (\ref{Leh}), can be calculated by,
\begin{eqnarray}
\bar L_{\rm eh}(\bm p,\omega)=\frac{-i}{2\pi}{\bar G}_{11}^<(\bm p,\omega),
\end{eqnarray}
where the lesser component of the single-particle Green's function $\hat G$ is defined as, 
\begin{eqnarray}
\hat {\bar G}^<(\bm p,\omega)
&=&
\frac{1}{2}[-\hat {\bar G}^{\rm R}(\bm p,\omega)+\hat {\bar G}^{\rm A}(\bm p,\omega)+\hat {\bar G}^{\rm K}(\bm p,\omega)].
\nonumber\\
\end{eqnarray}
The number of electrons or holes $n_{\rm eh}$ can be calculated as,
\begin{eqnarray}
n_{\rm eh}=\sum_{\bm p}\int_{-\infty}^\infty d\omega \bar L_{\rm eh}(\bm p,\omega).
\label{neh}
\end{eqnarray}
In addition, the number of (condensed) photons $n_{\rm ph}=|\avg{\bar{a_0}}|^2$ can also be obtained from Eq. (\ref{Langevin}) as, 
\begin{eqnarray}
&&n_{\rm ph}
=
\bigg|
-\frac{g}{\hbar\omega_{\rm cav}-2\mu-E_{\rm g}-i\kappa}
\sum_{\bm p}\avg{\bar c_{-\bm p,{\rm h}}\bar c_{\bm p,{\rm e}}}
\bigg|^2
\nonumber\\
&=&
\bigg|
i\frac{g}{\hbar\omega_{\rm cav}-2\mu-E_{\rm g}-i\kappa}
\sum_{\bm p}
\int_{-\infty}^\infty 
\frac{d\omega}{2\pi} \bar G_{12}^<(\bm p,\omega)
\bigg|^2.
\nonumber\\
\label{nph}
\end{eqnarray}

\begin{widetext}
\section{Derivation of Thouless criterion (\ref{Thouless})}\label{App_Thouless}

Here, we derive the Thouless criterion (\ref{Thouless}).
It is convenient to introduce
\begin{eqnarray}
\hat{\bar \Pi}^{\rm R}(\bm q,\omega)
&=&
-\frac{i}{2}\Big[
[\hat{\bar \Pi}(\bm q,\omega)]^{++}_{-+}
+\hat{\bar \Pi}(\bm q,\omega)]^{+-}_{--}
\Big],
\\
\hat{\bar \Pi}^{\rm A}(\bm q,\omega)
&=&
-\frac{i}{2}\Big[
[\hat{\bar \Pi}(\bm q,\omega)]^{+-}_{++}
+\hat{\bar \Pi}(\bm q,\omega)]^{--}_{-+}
\Big],
\\
\hat{\bar \Pi}^{\rm K}(\bm q,\omega)
&=&
-\frac{i}{2}\Big[
[\hat{\bar \Pi}(\bm q,\omega)]^{++}_{--}
+\hat{\bar \Pi}(\bm q,\omega)]^{--}_{++}
+\hat{\bar \Pi}(\bm q,\omega)]^{+-}_{-+}
\Big],
\end{eqnarray}
and
\begin{eqnarray}
\hat{\bar \Pi}^{\rm R}_{U}(\bm q,\omega)
&=&
-\frac{i}{2}\Big[
[\hat{\bar \Pi}_{U}(\bm q,\omega)]^{++}_{-+}
+\hat{\bar \Pi}_{U}(\bm q,\omega)]^{+-}_{--}
\Big],
\\
\hat{\bar \Pi}^{\rm A}_{U}(\bm q,\omega)
&=&
-\frac{i}{2}\Big[
[\hat{\bar \Pi}_{U}(\bm q,\omega)]^{+-}_{++}
+\hat{\bar \Pi}_{U}(\bm q,\omega)]^{--}_{-+}
\Big],
\\
\hat{\bar \Pi}^{\rm K}_{U}(\bm q,\omega)
&=&
-\frac{i}{2}\Big[
[\hat{\bar \Pi}_{U}(\bm q,\omega)]^{++}_{--}
+\hat{\bar \Pi}_{U}(\bm q,\omega)]^{--}_{++}
+\hat{\bar \Pi}_{U}(\bm q,\omega)]^{+-}_{-+}
\Big],
\end{eqnarray}
which enables us to rewrite Eq. (\ref{Sigph}) as 
\begin{eqnarray}
\hat{\bar \Sigma}^{\rm ph}_{\alpha,\beta}(\bm q,\omega)
&=&
-g^2
\left(
\begin{array}{cc}
\hat{\bar \Pi}_{U}^{\rm R}(\bm q,\omega) & \hat{\bar \Pi}_{U}^{\rm K}(\bm q,\omega) \\
0                                                    & \hat{\bar \Pi}_{U}^{\rm A}(\bm q,\omega) 
\end{array}
\right)_{\alpha,\beta}
+
\left(
\begin{array}{cc}
-i\kappa & -2i\kappa\tau_3 \\
0           & i\kappa 
\end{array}
\right)_{\alpha,\beta},
\end{eqnarray}
and Eq. (\ref{PiU_index}) as,
\begin{eqnarray}
&&\left(
\begin{array}{cc}
\hat{\bar \Pi}_{U}^{\rm R}(\bm q,\omega) & \hat{\bar \Pi}_{U}^{\rm K}(\bm q,\omega) \\
0                                                    & \hat{\bar \Pi}_{U}^{\rm A}(\bm q,\omega) 
\end{array}
\right)
=
\left(
\begin{array}{cc}
\hat{\bar \Pi}^{\rm R}(\bm q,\omega) & \hat{\bar \Pi}^{\rm K}(\bm q,\omega) \\
0                                               & \hat{\bar \Pi}^{\rm A}(\bm q,\omega) 
\end{array}
\right)
+U
\left(
\begin{array}{cc}
\hat{\bar \Pi}^{\rm R}(\bm q,\omega) & \hat{\bar \Pi}^{\rm K}(\bm q,\omega) \\
0                                                & \hat{\bar \Pi}^{\rm A}(\bm q,\omega) 
\end{array}
\right)
\left(
\begin{array}{cc}
\hat{\bar \Pi}_{U}^{\rm R}(\bm q,\omega) & \hat{\bar \Pi}_{U}^{\rm K}(\bm q,\omega) \\
0                                                    & \hat{\bar \Pi}_{U}^{\rm A}(\bm q,\omega) 
\end{array}
\right).
\label{PiU_RAK}
\end{eqnarray}
From Eq. (\ref{PiU_RAK}), we obtain the retarded component of $\hat{\bar\Pi}_U$,
\begin{eqnarray}
\hat{\bar \Pi}_U^{\rm R}(\bm q,\omega)
=[1-U\hat{\bar \Pi}^{\rm R}(\bm q,\omega)]^{-1}\hat{\bar \Pi}^{\rm R}(\bm q,\omega),
\end{eqnarray}
as well as the photon Green's function,
\begin{eqnarray}
[\hat{\bar D}^{\rm R}(\bm q,\omega)]^{-1}
&=&[\hat{\bar D}^{0\rm R}(\bm q,\omega)]^{-1}
-\hat{\bar \Sigma}_{\rm ph}^{\rm R}(\bm q,\omega)
=[\hat {\bar D}_{\rm neq}^{0\rm R}(\bm q,\omega)]^{-1}
+g^2 \Pi_U^{\rm R}(\bm q,\omega) 
\nonumber\\
&=&[1-U\hat{\bar\Pi}^{\rm R}(\bm q,\omega)]^{-1}
\big[[U+g^2 \hat{\bar D}_{\rm neq}^{0\rm R}(\bm q,\omega)]^{-1}
-\hat{\bar\Pi}^{\rm R}(\bm q,\omega)
\big]
[U+g^2 \hat{\bar D}_{\rm neq}^{0\rm R}(\bm q,\omega)]
[\hat{\bar D}_{\rm neq}^{0\rm R}(\bm q,\omega)]^{-1},
\label{DR_App}
\end{eqnarray}
where we have introduced
\begin{eqnarray}
[\hat {\bar D}_{\rm neq}^{0\rm R}(\bm q,\omega)]^{-1}
=[\hat {\bar D}^{0\rm R}(\bm q,\omega)]^{-1}+i\kappa.
\end{eqnarray}
Taking the determinant of $\hat{\bar D}^{\rm R}$ at $\bm q=\omega=0$, we obtain  
\begin{eqnarray}
&&{\rm det}[\hat{\bar D}^{\rm R}(\bm 0,0)]^{-1}
=
{\rm det}[1-U\hat{\bar\Pi}^{\rm R}(\bm 0,0)]^{-1}
{\rm det}
\big[[U+g^2 \hat{\bar D}_{\rm neq}^{0\rm R}(\bm 0,0)]^{-1}
-\hat{\bar\Pi}^{\rm R}(\bm 0,0)
\big]
{\rm det}[U+g^2 \hat{\bar D}_{\rm neq}^{0\rm R}(\bm 0,0)]
{\rm det}[\hat{\bar D}_{\rm neq}^{0\rm R}(\bm 0,0)]^{-1}
\nonumber\\
&&=
{\rm det}[1-U\hat{\bar\Pi}^{\rm R}(\bm 0,0)]^{-1}
{\rm det}\left(
\begin{array}{cc}
\frac{1}{U_{\rm eff}}-{\bar\Pi}_{+-}^{\rm R}(\bm 0,0) & -{\bar\Pi}_{++} ^{\rm R}(\bm 0,0) \\
-{\bar\Pi}_{--}^{\rm R}(\bm 0,0) & \frac{1}{U_{\rm eff}^*}-{\bar\Pi}_{-+}^{\rm R}(\bm 0,0)
\end{array}
\right)
{\rm det}[U+g^2 \hat{\bar D}_{\rm neq}^{0\rm R}(\bm 0,0)]
{\rm det}[\hat{\bar D}_{\rm neq}^{0\rm R}(\bm 0,0)]^{-1}
\nonumber\\
&&=
{\rm det}[1-U\hat{\bar\Pi}^{\rm R}(\bm 0,0)]^{-1}
\frac{1}{4}
{\rm det}\left(
\begin{array}{cc}
{\rm Re}\big[\frac{2}{U_{\rm eff}}\big]-{\bar\Xi}_{11}^{\rm R}(\bm 0,0) 
& {\rm Im}\big[\frac{2}{U_{\rm eff}}\big]-{\bar\Xi}_{12}^{\rm R}(\bm 0,0) \\
{\rm Im}\big[\frac{2}{U_{\rm eff}}\big]-{\bar\Xi}_{21}^{\rm R}(\bm 0,0)
& {\rm Re}\big[\frac{2}{U_{\rm eff}}\big]-{\bar\Xi}_{22}^{\rm R}(\bm 0,0)
\end{array}
\right)
{\rm det}[U+g^2 \hat{\bar D}_{\rm neq}^{0\rm R}(\bm 0,0)]
{\rm det}[\hat{\bar D}_{\rm neq}^{0\rm R}(\bm 0,0)]^{-1},
\nonumber\\
\label{Thouless_2}
\end{eqnarray}
\end{widetext}
where $\hat{\bar\Xi}^{\rm R}=2\hat\Lambda\hat{\bar\Pi}^{\rm R}\Lambda^{-1}$ with 
\begin{eqnarray}
\hat\Lambda
=\frac{1}{\sqrt 2}
\left(
\begin{array}{cc}
1 & 1 \\
i & -i
\end{array}
\right),
\end{eqnarray}
is the lowest-order pair-correlation transformed to the amplitude-phase representation \cite{Ohashi2003}.
Recalling 
\begin{eqnarray}
\hat{\bar\Pi}^{\rm R}(\bm 0,0)
=
\left(
\begin{array}{cc}
{\bar \Pi}_{-+}^{\rm R}(\bm 0,0) & {\bar \Pi}_{--}^{\rm R}\bm 0,0) \\
{\bar \Pi}_{++}^{\rm R}(\bm 0,0) & {\bar \Pi}_{+-}^{\rm R}(\bm 0,0) 
\end{array}
\right),
\end{eqnarray}
and ($s,s'=+,-$)
\begin{eqnarray}
{\bar \Pi}_{s,s'}^{\rm R}(\bm 0,0) 
&=&\frac{i}{2}\sum_{\bm p}
\int_{-\infty}^\infty \frac{d\omega}{2\pi}
{\rm Tr}[\tau_s \bar G^{\rm R}(\bm p,\omega)\tau_{s'} \bar G^{\rm K}(\bm p,\omega)
\nonumber\\
&+&
\tau_s \bar G^{\rm K}(\bm p,\omega)\tau_{s'} \bar G^{\rm A}(\bm p,\omega)],
\end{eqnarray}
we can derive the following relations \cite{Hanai2017},
\begin{eqnarray}
\bar\Xi_{22}^{\rm R}(\bm 0,0)
&=&\frac{i}{2}\sum_{\bm p}
\int_{-\infty}^\infty \frac{d\omega}{2\pi}
{\rm Tr}[\tau_2 \hat{\bar G}^{\rm R}(\bm p,\omega)\tau_2 \hat{\bar G}^{\rm K}(\bm p,\omega)
\nonumber\\
&&+
\tau_2 \hat{\bar G}^{\rm K}(\bm p,\omega)\tau_2 \hat{\bar G}^{\rm A}(\bm p,\omega)]
\nonumber\\
&=&
\sum_{\bm p}
\int_{-\infty}^\infty \frac{d\omega}{2\pi}
{\rm Re}\bigg[
-i\frac{\bar G_{12}^{\rm K}(\bm p,\omega)}{\Delta_0}
\bigg],
\\
\bar\Xi_{12}^{\rm R}(\bm 0,0)
&=&\frac{i}{2}\sum_{\bm p}
\int_{-\infty}^\infty \frac{d\omega}{2\pi}
{\rm Tr}[\tau_1 \hat{\bar G}^{\rm R}(\bm p,\omega)\tau_2 \hat{\bar G}^{\rm K}(\bm p,\omega)
\nonumber\\
&&+\tau_1 \hat{\bar G}^{\rm K}(\bm p,\omega)\tau_2 \hat{\bar G}^{\rm A}(\bm p,\omega)]
\nonumber\\
&=&
\sum_{\bm p}
\int_{-\infty}^\infty \frac{d\omega}{2\pi}
{\rm Im}\bigg[
-i\frac{\bar G_{12}^{\rm K}(\bm p,\omega)}{\Delta_0}
\bigg].
\end{eqnarray}
By using the above relations and the nonequilibrium steady state gap equation (\ref{GAP_Gk}), we get
\begin{eqnarray}
{\rm Re}\bigg[\frac{2}{U_{\rm eff}}\bigg]-{\bar\Xi}_{22}^{\rm R}(\bm 0,0)=0, \\
{\rm Im}\bigg[\frac{2}{U_{\rm eff}}\bigg]-{\bar\Xi}_{12}^{\rm R}(\bm 0,0)=0.
\end{eqnarray}
Substituting these into Eq. (\ref{Thouless_2}) yields the desired Thouless criterion (\ref{Thouless}).


\section{Inter- and intra-band contributions to $\hat\Pi$}\label{App_interintra}

Here, we split the lowest-order pair-correlation function $\hat\Pi$ into the inter- ($\hat \Pi_{\rm inter}$) and intra-band ($\hat \Pi_{\rm intra}$) contribution. 
We first split the single-particle Green's function $\hat G=\hat G_{\rm l}+\hat G_{\rm u}$ into the lower ($\hat G_{\rm l}$) and upper ($\hat G_{\rm u}$) branch contribution.
The retarded component $\hat G^{\rm R}=\hat G^{\rm R}_{\rm l}+\hat G^{\rm R}_{\rm u}$ is split as,
\begin{eqnarray}
\hat{\bar G}^{\rm R}_{\rm l}(\bm p,\omega)
&=&\frac{1}{2}
\bigg[
\bm 1-\frac{\xi_{\bm p}}{E_{\bm p}}\tau_3-\frac{\Delta_0}{E_{\bm p}}\tau_1
\bigg]
\frac{1}{\hbar\omega+i\gamma+E_{\bm p}},
\nonumber\\
\\
\hat{\bar G}^{\rm R}_{\rm u}(\bm p,\omega)
&=&\frac{1}{2}
\bigg[
\bm 1+\frac{\xi_{\bm p}}{E_{\bm p}}\tau_3+\frac{\Delta_0}{E_{\bm p}}\tau_1
\bigg]
\frac{1}{\hbar\omega+i\gamma-E_{\bm p}},
\nonumber\\
\end{eqnarray}
where $\hat G_{\rm l(u)}^{\rm R}$ has a pole at the lower (upper) branch, $\hbar\omega=-E_{\bm p}-i\gamma (\hbar\omega=E_{\bm p}-i\gamma)$.
The advanced component of the lower (upper) contribution is given by $\hat G^{\rm A}_{\rm l(u)}=[\hat G^{\rm R}_{\rm l(u)}]^\dagger$.
We also split the Keldysh component $\hat G^{\rm K}=\hat G^{\rm K}_{\rm l}+\hat G^{\rm K}_{\rm u}$ to the lower ($\hat G^{\rm K}_{\rm l}$) and upper ($\hat G^{\rm K}_{\rm u}$) contributions, by rewritting $\hat G^{\rm K}$ in the form
\begin{eqnarray}
\hat{\bar G}^{\rm K}(\bm p,\omega)
&=&
\hat{\bar G}^{\rm R}(\bm p,\omega)\hat F(\bm p,\omega) 
-\hat F(\bm p,\omega) \hat{\bar G}^{\rm A}(\bm p,\omega)
\nonumber\\
&=&
\hat G^{\rm K}_{\rm l}+\hat G^{\rm K}_{\rm u},
\label{Gksplit}
\end{eqnarray}
where
\begin{eqnarray}
\hat{\bar G}^{\rm K}_{\rm l}(\bm p,\omega)
&=&
\hat{\bar G}^{\rm R}_{\rm l}(\bm p,\omega)\hat F(\bm p,\omega) 
-\hat F(\bm p,\omega) \hat{\bar G}^{\rm A}_{\rm l}(\bm p,\omega),
\nonumber\\
\\
\hat{\bar G}^{\rm K}_{\rm u}(\bm p,\omega)
&=&
\hat{\bar G}^{\rm R}_{\rm u}(\bm p,\omega)\hat F(\bm p,\omega) 
-\hat F(\bm p,\omega) \hat{\bar G}^{\rm A}_{\rm u}(\bm p,\omega).
\nonumber\\
\end{eqnarray}
In obtaining the first equality of Eq. (\ref{Gksplit}),
we have introduced the Hermitian matrix $\hat F$ given by,
\begin{widetext}
\begin{eqnarray}
&&\hat F(\bm p,\omega)
=
F_-(\omega)\bm 1
+\frac{\xi_{\bm p}^2+\gamma^2}{E_{\bm p}^2+\gamma^2} F_+(\omega)\tau_3
+\frac{\Delta_0\xi_{\bm p}}{E_{\bm p}^2+\gamma^2}F_+(\omega) \tau_1
+\frac{\Delta_0\gamma}{E_{\bm p}^2+\gamma^2}F_+(\omega) \tau_2,
\end{eqnarray} 
where we have used the fact that $\hat G^{\rm K}$ is anti-Hermitian ($\hat G^{\rm K}=-[\hat G^{\rm K}]^\dagger$) and $\hat G^{\rm A}=[\hat G^{\rm R}]^\dagger$.

Using these definitions of the lower ($\hat G_{\rm l}$) and upper ($\hat G_{\rm u}$) contribution of the single-particle Green's function, we define the inter- ($\hat \Pi_{\rm inter}$) and intra-band ($\hat \Pi_{\rm intra}$) contribution of $\hat \Pi=\hat \Pi_{\rm inter}+\hat \Pi_{\rm intra}$ as, 
\begin{eqnarray}
[{\bar \Pi}_{s,s'}^{\rm inter}(\bm q,\omega)]^{\alpha,\beta}_{\alpha',\beta'}
&=&
-\sum_{\bm k}\int_{-\infty}^{\infty}\frac{\hbar d\omega_1}{2\pi}
{\rm Tr} \Big[
\tau_s \hat{\bar G}_{\alpha,\beta}^{\rm l}(\bm k+\frac{\bm q}{2},\omega_1+\frac{\omega}{2}) \tau_{s'}  \hat{\bar G}_{\beta',\alpha'}^{\rm u}(\bm k-\frac{\bm q}{2},\omega_1-\frac{\omega}{2})
\nonumber\\
&+&
\tau_s \hat{\bar G}_{\alpha,\beta}^{\rm u}(\bm k+\frac{\bm q}{2},\omega_1+\frac{\omega}{2}) \tau_{s'}  \hat{\bar G}_{\beta',\alpha'}^{\rm l}(\bm k-\frac{\bm q}{2},\omega_1-\frac{\omega}{2})
\Big],
\label{PIinter}
\\
{}[{\bar \Pi}_{s,s'}^{\rm intra}(\bm q,\omega)]^{\alpha,\beta}_{\alpha',\beta'}
&=&
-\sum_{\bm k}\int_{-\infty}^{\infty}\frac{\hbar d\omega_1}{2\pi}
{\rm Tr} \Big[
\tau_s \hat{\bar G}_{\alpha,\beta}^{\rm l}(\bm k+\frac{\bm q}{2},\omega_1+\frac{\omega}{2}) \tau_{s'}  \hat{\bar G}_{\beta',\alpha'}^{\rm l}(\bm k-\frac{\bm q}{2},\omega_1-\frac{\omega}{2})
\nonumber\\
&+&
\tau_s \hat{\bar G}_{\alpha,\beta}^{\rm u}(\bm k+\frac{\bm q}{2},\omega_1+\frac{\omega}{2}) \tau_{s'}  \hat{\bar G}_{\beta',\alpha'}^{\rm u}(\bm k-\frac{\bm q}{2},\omega_1-\frac{\omega}{2})
\Big].
\label{PIintra}
\end{eqnarray}
\end{widetext}

\end{appendix}
\par

\end{document}